\documentclass[twocolumn, tighten, times]{aastex631}
\usepackage{epsfig}

\newcommand{\qpah}{$\mathrm{q}_{\mathrm{PAH}}$}

\newcommand{\av}{\ensuremath{A(V)}}
\newcommand{\avsmc}{\ensuremath{A(V)_\mathrm{SMC}}}
\newcommand{\rv}{\ensuremath{R(V)}}
\newcommand{\rvsmc}{\ensuremath{R(V)_\mathrm{SMC}}}
\newcommand{\ebv}{\ensuremath{E(B-V)}}
\newcommand{\ebvmw}{\ensuremath{E(B-V)_\mathrm{MW}}}
\newcommand{\ebvsmc}{\ensuremath{E(B-V)_\mathrm{SMC}}}
\newcommand{\nhismc}{\ensuremath{N(HI)_\mathrm{SMC}}}
\newcommand{\nhi}{\ensuremath{N(HI)}}

\newcommand{\nhiav}{\ensuremath{N(HI)/A(V)}}
\newcommand{\alav}{\ensuremath{A(\lambda)/A(V)}}
\newcommand{\elvebv}{\ensuremath{E(\lambda - V)/E(B - V)}}
\newcommand{\fbump}{2175~\AA}

\newcommand{\teff}{\ensuremath{T_{\mathrm{eff}}}}
\newcommand{\logg}{\ensuremath{\log g}}
\newcommand{\abund}{\ensuremath{[\mathrm{m/H}]}}
\newcommand{\vturb}{\ensuremath{v_{\mathrm{turb}}}}
\newcommand{\vrot}{\ensuremath{v \sin i}}
\newcommand{\vrad}{\ensuremath{v_{\mathrm{rad}}}}
\newcommand{\kms}{\ensuremath{\mathrm{km~s}^{-1}}}

\newcommand{\kchange}[1]{{#1}}

\begin{document}

\shortauthors{Gordon et al.}
\shorttitle{SMC 2175~\AA\ Extinction}

\title{Expanded Sample of Small Magellanic Cloud Ultraviolet Dust Extinction Curves: \\ Correlations between the \fbump\ bump, \qpah, UV extinction shape, and \nhiav}

\author[0000-0001-5340-6774]{Karl~D.~Gordon}
\affiliation{Space Telescope Science Institute, 3700 San Martin
  Drive, Baltimore, MD, 21218, USA}
\affiliation{Sterrenkundig Observatorium, Universiteit Gent,
  Gent, Belgium}

\author[0000-0002-2371-5477]{E.\ L.\ Fitzpatrick}
\affiliation{Department of Astronomy \& Astrophysics, Villanova University, 800 Lancaster Avenue, Villanova, PA 19085, USA}

\author[0000-0002-9139-2964]{Derck Massa}
\affil{Space Science Institute, 4750 Walnut Street, Suite 205, Boulder, CO 80301, USA}

\author[0000-0001-9806-0551]{Ralph Bohlin}
\affiliation{Space Telescope Science Institute, 3700 San Martin
  Drive, Baltimore, MD, 21218, USA}

\author[0000-0002-5235-5589]{J\'er\'emy Chastenet}
\affiliation{Sterrenkundig Observatorium, Universiteit Gent,
  Gent, Belgium}

\author[0000-0002-7743-8129]{Claire E. Murray}
\affiliation{Space Telescope Science Institute, 3700 San Martin Drive, Baltimore, MD, 21218, USA}
\affiliation{Department of Physics \& Astronomy, Johns Hopkins University, 3400 N. Charles Street, Baltimore, MD 21218}

\author[0000-0002-0141-7436]{Geoffrey~C.~Clayton}
\affiliation{Department of Physics \& Astronomy, Louisiana State University, Baton Rouge, LA 70803, USA}

\author[0000-0003-3063-4867]{Daniel J. Lennon}
\affiliation{Instituto de Astrofísica de Canarias, 38 200, La Laguna, Tenerife, Spain} 
\affiliation{Dpto. Astrofísica, Universidad de La Laguna, 38 205, La Laguna, Tenerife, Spain}

\author{Karl A.\ Misselt}
\affiliation{Steward Observatory, University of Arizona, Tucson,
  AZ 85721, USA}

\author[0000-0002-4378-8534]{Karin~Sandstrom}
\affiliation{Department of Astronomy \& Astrophysics, University of California, San Diego, 9500 Gilman Drive, San Diego, CA 92093, USA}

\begin{abstract} 
The Small Magellanic Cloud (SMC) shows a large variation in ultraviolet (UV) dust extinction curves, ranging from Milky Way-like (MW) to significantly steeper curves with no detectable 2175~\AA\ bump.
This result is based on a sample of only nine sightlines.
From HST/STIS and IUE spectra of OB stars, we have measured UV extinction curves along 32 SMC sightlines where eight of these curves were published previously.
We find 16 sightlines with steep extinction with no detectable 2175~\AA\ bump, four sightlines with MW-like extinction with a detectable 2175~\AA\ bump, two sightlines with fairly flat UV extinction and weak/absent 2175~\AA\ bumps, and 10 sightlines with unreliable curves due to low SMC dust columns.
Our expanded sample shows that the sightlines with and without the 2175~\AA\ bump are located throughout the SMC and not limited to specific regions.
The average extinction curve of the 16 bumpless sightlines is very similar to the previous average based on four sightlines.
We find no correlation between dust column and the strength of the 2175~\AA\ bump.
We test the hypothesis that the 2175~\AA\ bump is due to the same dust grains that are responsible for the mid-infrared carbonaceous (PAH) emission features and find they are correlated, confirming recent work in the MW.
Overall, the slope of the UV extinction increases as the amplitudes of the 2175~\AA\ bump and far-UV curvature decrease.
Finally, the UV slope is correlated with $N(HI)/A(V)$ and the 2175~\AA\ bump and nonlinear far-UV rise amplitudes are anti-correlated with $N(HI)/A(V)$.
\end{abstract}


\section{Introduction}
\label{sec_intro}

Dust grains modulate the flow of radiation through a galaxy, affecting the general ISM structure, the environments of star formation, and the overall energy balance.
In addition, dust complicates our ability to understand the properties of the stellar populations and interstellar gas. 
Characterizing the wavelength-dependence of dust extinction can enable us to correct for its effects and provide observational constraints on dust grain size distributions.

Investigating the dust extinction in external galaxies, in conjunction with Milky Way (MW) studies, is important because it increases the range over which quantifiable environmental factors (e.g., metallicity, gas-to-dust ratio, IR intensities, H$\alpha$ intensity, etc.) can be correlated with extinction properties and the properties of the grain populations themselves (e.g., grain size, shape, and composition). 
This, in turn, increases our understanding of the sensitivity of dust grains to environmental factors and also leads to a measure of predictability - i.e., certain environmental factors may yield certain extinction properties - which is useful for correcting for dust's effects on observations of stars and gas. 

Ultraviolet extinction curves have been determined in the Local Group of galaxies for the MW, the Large Magellanic Cloud (LMC), the Small Magellanic Cloud (SMC), and
M31.
The extinction curves in these galaxies paint a complex picture of the environmental dependence of dust properties.
In the MW, the near-infrared to UV extinction curves can be described fairly well by a relationship that depends on only one parameter, $\rv = \av/\ebv$, which is a measure of the overall dust grain size \citep{Cardelli89, Valencic04, Fitzpatrick19, Gordon23}.
There are significant deviations from the \rv\ relationship in different Galactic environments \citep{Mathis92, Valencic04} including one sightline that (after foreground extinction is removed) is similar to the traditional SMC curve \citep{Valencic03}.
In the LMC, the UV extinction curves show a distinctly different behavior between the LMC~2 (roughly 30~Dor region) and the rest of the LMC \citep{Fitzpatrick85, Fitzpatrick86, Misselt99}.
The 2175~\AA\ bump is much weaker in the LMC~2 region, whereas the rest of the LMC shows curves similar to the average Galactic extinction curve.
The SMC has two kinds of curves: the traditional SMC curve without a 2175~\AA\ bump and those with a recognizable 2175~\AA\ bump \citep{Gordon03, MaizApellaniz12}.
In M31, recent work has found two extinction curves consistent with the MW average and two that are similar to the LMC~2 average \citep{Clayton15}.
The extinction curve behavior in these four galaxies implies that the physical properties of dust grains are likely to be dependent on a multitude of environmental parameters.
These include the gas-to-dust ratio, metallicity, and star formation activity, all of which may affect the overall composition and size distribution of dust grains \citep{Gordon97, Gordon03}.

In recent years, observations of UV extinction in distant galaxies have advanced from, first, noting the absence of the 2175~\AA\ bump in many starburst and Lyman Break galaxies \citep{Calzetti94, Gordon97, Vijh03} to more recent detections of the bump in the spectra of high redshift galaxies and an anti-correlation between its strength and galaxy luminosity 
\citep[e.g.,][]{Noll07, Noll09, Buat12, Kriek13, Witstok23, Markov23} and the discovery of strong bumps in the dust extinction of high-redshift gamma ray bursts \citep[e.g.,][]{Eliasdottir09, Schady12, Zafar12}.
The emerging picture of dust extinction properties in distant galaxies is not unlike that seen in the Local Group, i.e., extreme regional variations suggesting significant internal variations in the basic properties of the interstellar grain populations.
In addition, strong galaxy-to-galaxy trends suggest that global properties, such as metallicity and recent star formation activity, may also play a significant role in shaping the dust and resultant extinction properties.

The SMC is a critical galaxy in which to study dust properties.
It has been known for many years to have UV extinction curves that differ the most from typical MW extinction \citep{Lequeux82, Prevot84} suggesting systematically different grain properties.
It is also an ideal laboratory for investigating the nature of the 2175~\AA\ bump carrier, as the SMC shows the largest range of 2175~\AA\ bump strengths of any Local Group galaxy, indicating large internal variations in grain properties. 
In addition, it has a significantly lower metallicity than the MW \citep{Russell92, Dominguez-Guzman22} and such a chemically primitive environment may be more similar to that in more distant galaxies (e.g., gamma-ray burst host galaxies). 
Thus, the SMC could provide an important link between the dust grain and extinction properties in the nearby universe with those of the observationally less accessible distant galaxies.

Given its importance, it is somewhat surprising that the published sample of UV extinction curves for the SMC is small, with a total of only 9 sightlines.
This sample is composed of four sightlines created from {\it International Ultraviolet Explorer} spectroscopic observations \citep{Gordon98}, one sightline created from {\it Hubble Space Telescope} (HST) {\it Space Telescope Imaging Spectrograph} (STIS) slit spectroscopy observations \citep{Gordon03}, and four sightlines created by fitting HST/STIS slitless prism spectroscopy observations \citep{MaizApellaniz12}.  
These show large differences in UV extinction strengths, especially in the 2175~\AA\ bump and far-UV rise features.
Roughly, these nine curves can be categorized into three that are are similar to those seen in the MW with recognizable 2175~\AA\ bumps (AzV~456, MR12~09, 10) and six that are roughly linear with $\lambda^{-1}$ and lack a 2175~\AA\ bump (AzV~18, 23, 214, 398, MR12~08, 11).
Based on the 5 sightlines known at the time, \citet{Gordon03} postulated that the differences could be due to processing from nearby star formation since all the sightlines without 2175~\AA\ bumps were in the SMC's star-forming Bar and the remaining sightline with a 2175~\AA\ bump was in the more quiescent SMC Wing.  
This interpretation was clearly too simple as the 4 sightlines studied by \citet{MaizApellaniz12} were all in a small region in the SMC's star-forming Bar and split evenly between those with and without a 2175~\AA\ bump.
Clearly, the small sample of measured UV extinction curves is the major limitation in studying the UV extinction properties in the SMC.

In this paper, we present UV through near-IR extinction curves for the sightlines towards 32 stars, greatly enhancing the previously available sample.
In the past, one of the main deterrents to measuring SMC extinction curves was the lack of large numbers of high quality stellar spectral types, making it difficult to identify reddened stars.
In recent years, however, the number of stars with good spectral types has increased dramatically, largely due to the 2dF survey of more than 4,000 SMC stars \citep{Evans04}, plus a number of VLT-FLAMES surveys \citep{Evans06, Martayan07}.
This better characterization of the SMC's hot star population -- from which multiband extinction curves are derived -- provided a main motivation for this work. 

In \S\ref{sec_data} we describe new {\it HST/STIS} UV spectrophotometric observations for 19 SMC stars, along with the data processing techniques.
This section also gives the details of the archival {\it IUE} and {\it HST/STIS} spectra used here and ancillary data for the entire sample.
In \S\ref{sec_ext_curves}, we present the calculation of the extinction curves, the correction for MW foreground extinction, the best-fitting stellar and extinction parameters, and a comparison to previous work.
The UV extinction behavior distribution across the SMC, sample average curves, and correlations with mid-IR carbonaceous (aka PAH) features are presented in \S\ref{sec_discussion}.
The behavior of the SMC UV extinction in context with results for the LMC and MW are analyzed in \S\ref{sec_discussion2} focusing on the general sightline characteristics, the correlations between UV extinction parameters, and correlations with gas-to-dust.
Finally, the conclusions of this work are given in \S\ref{sec_conclusions}.

\section{Data}
\label{sec_data}

\begin{deluxetable}{lllcc}
\tabletypesize{\footnotesize}
\tablecaption{Sample Details\label{tab_sample}}
\tablehead{\colhead{Name} & \colhead{RA (J2000)\tablenotemark{a}} & 
   \colhead{DEC (J2000)\tablenotemark{a}} & \colhead{SpType} & \colhead{Ref} }
\startdata
\multicolumn{5}{c}{STIS Sample} \\ \hline
2dFS 413    & 00 42 15.81 & -73 24 32.9 & B1-5(IV) & 1 \\
2dFS 626    & 00 46 55.80 & -73 06 14.6 & B0-5(V)  & 1 \\
2dFS 662    & 00 47 45.58 & -73 13 37.2 & B1-2(III) & 1 \\
2dFS 699    & 00 47 58.49 & -72 31 41.8 & B0.5(II)e & 1 \\
2dFS 3014    & 01 14 26.28 & -73 17 13.5 & B0-5(V) & 1 \\
2dFS 3030    & 01 14 40.14 & -73 16 14.5 & O7-8V   & 1 \\
2dFS 3171    & 01 16 09.19 & -73 12 38.8 & B0(V)   & 1 \\ 
AzV 4        & 00 45 03.30 & -72 41 57.4 & B2Ib    & 2 \\
AzV 23\tablenotemark{b} & 00 47 38.91 & -73 22 53.9 & B3Ia    & 3 \\
AzV 86       & 00 51 05.83 & -72 00 30.6 & B1Ia    & 3 \\
AzV 132      & 00 52 51.24 & -73 06 53.6 & B2      & 4 \\
AzV 218      & 00 59 04.34 & -72 19 40.8 & B2Iab   & 2 \\
AzV 404\tablenotemark{d} & 01 06 29.26 & -72 22 08.6 & B2.5Iab & 3 \\  
MR12 06/07\tablenotemark{d} & 00 45 06.51 & -73 18 24.6 & B6-7 & 5 \\  
MR12 09\tablenotemark{c} & 00 45 35.10 & -73 18 36.0 & B2.5-5  & 5 \\
MR12 10\tablenotemark{c} & 00 45 37.10 & -73 18 40.0 & B3-6    & 5 \\
MR12 11\tablenotemark{c} & 00 45 34.49 & -73 18 41.8 & B1.5-3  & 5 \\  
NGC330 ELS 110 & 00 56 20.67 & -72 26 25.5 & B2IV  & 6 \\
NGC330 ELS 114 & 00 56 57.0 & -72 25 31.7 & B2III  & 6 \\
NGC330 ELS 116 & 00 55 26.5 & -72 27 33.7 & B3III  & 6 \\
NGC346 ELS 056\tablenotemark{d} & 00 58 56.1 & -72 09 33.8 & B0V    & 6 \\  
SMC5-398     & 00 54 02.67 & -72 25 40.8 & B0.5III & 7 \\
SMC5-3739    & 00 57 13.85 & -72 22 30.1 & O9III   & 7 \\
SMC5-79264   & 00 55 50.04 & -72 19 23.7 & B3III   & 7 \\ 
SMC5-82923   & 00 53 26.17 & -72 11 40.0 & B0III   & 7 \\ \hline
\multicolumn{5}{c}{IUE Sample} \\ \hline
AzV 18\tablenotemark{b} & 00 47 12.22 & -73 06 33.1 & B3Ia     & 2 \\
AzV 70      & 00 50 18.12 & -72 38 10.0 & O9Ia     & 2 \\
AzV 214\tablenotemark{b} & 00 58 54.78 & -72 13 17.2 & B2Ia     & 2 \\
AzV 289     & 01 01 58.78 & -72 35 39.0 & B0Ia     & 8 \\
AzV 380     & 01 05 24.76 & -73 03 53.0 & B1Ia:    & 2 \\
AzV 398\tablenotemark{b} & 01 06 09.81 & -71 56 00.7 & O9Ia:    & 2 \\
AzV 456\tablenotemark{b} & 01 10 55.72 & -72 42 56.1 & O8II     & 2 \\
AzV 462     & 01 11 25.92 & -72 31 21.3 & B2Ia     & 2 \\ 
BBB SMC280     & 00 48 05.15 & -73 12 03.6 & B0-1(I)  & 9 \\
SK 191      & 01 41 42.07 & -73 50 38.2 & B1.5Ia   & 3 \\
\enddata
\tablenotetext{a}{Coordinates are from GAIA DR2 \citep{2018yCat.1345....0G} except for the NGC330/NGC346 sources \citep{Evans06}.}
\tablenotetext{b}{Extinction curves previously published by \citet{Gordon98} and \citet{Gordon03}.}
\tablenotetext{c}{Extinction curves based on STIS NUV prism data previously published by \citet{MaizApellaniz12}.}
\tablenotetext{d}{Extinction curve could not be computed as the sightline \ebv\ values were too low.  Included for completeness as the STIS spectra were taken as part of the HST extinction curve programs.}
\tablerefs{(1) \citet{Evans04}, (2) \citet{SmithNeubig97}, (3) \citet{Lennon97}, (4) \citet{Azzopardi75}, (5) \citet{MaizApellaniz12}, (6) \citet{Evans06}, (7) \citet{Martayan07}, (8) \citet{Walborn83}, and (9) \citet{Nandy82}.}
\end{deluxetable}

\begin{figure*}[tbp]
\epsscale{1.15}
\plotone{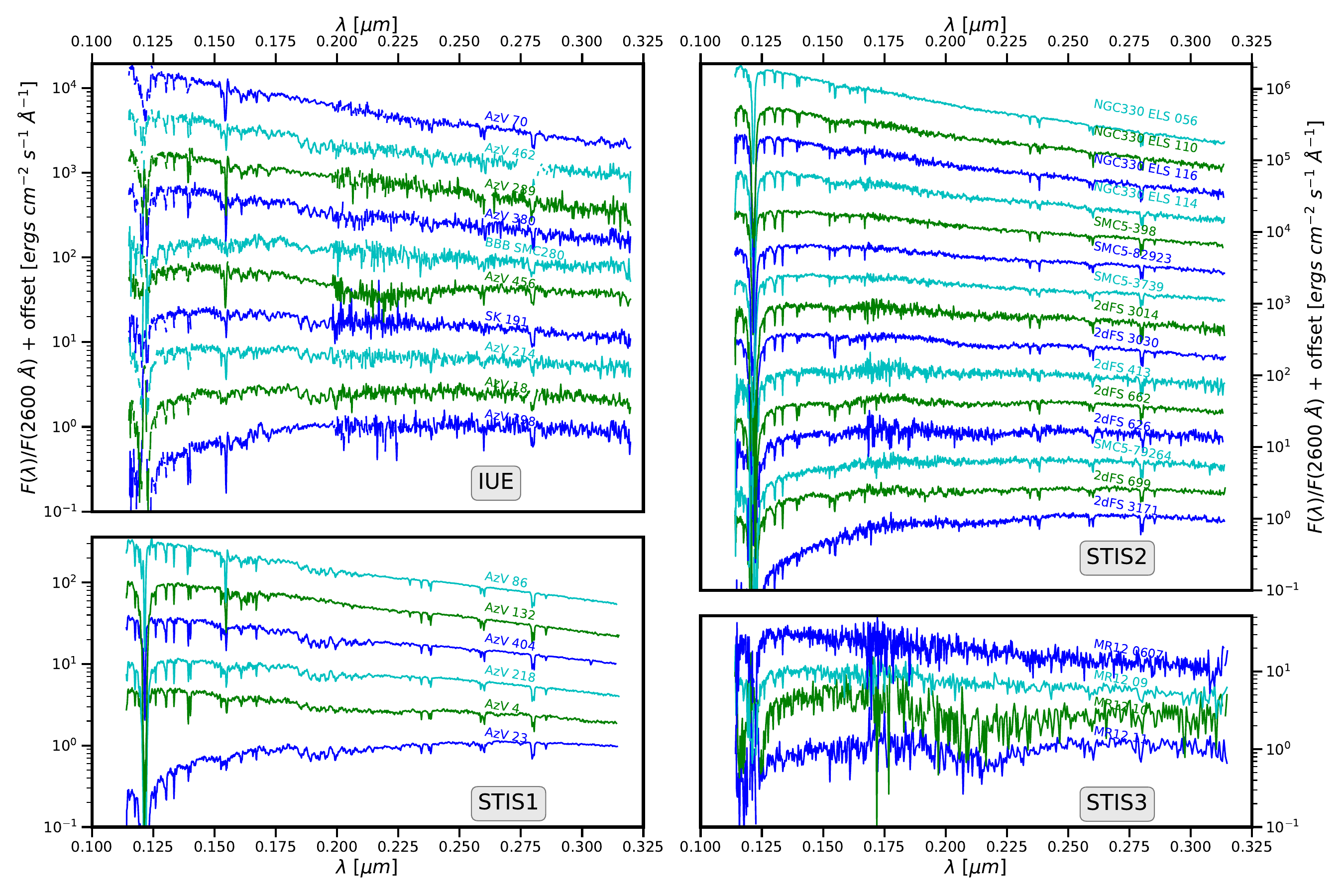}
\caption{The UV spectra for all the stars in our sample. 
In each panel, the spectra are in order of UV spectral slope.
This sorting shows all the spectra with little overlap and provides a rough ordering by amount of dust extinction.
The IUE panel depicts all the IUE archival spectra; and the STIS1, STIS2, and STIS3 panels show the spectra from HST programs 8198, 12258, and 14225, respectively.
  \label{fig_uv_spectra}}
\end{figure*}

The sample of extinction curve sightlines is composed of all OB stars with spectroscopic observations covering the UV region from 1150 to 3100~\AA\ with with IUE or HST/STIS slit spectral observations.
The STIS sample is composed of all the targets in our three HST programs on this subject (PIDs: 8198, 12258, and 14225; PI: Gordon).  
The targets in the first program were used to create extinction curves using the standard pair method and published by \citet{Gordon03}.  
The targets in the second and third programs are presented here for the first time.
The IUE sample is composed principally of the reddened and comparison stars from \citet{Gordon98} supplemented with two additional stars (SK~191 and BBB~SMC~280) found by a careful followup search of the IUE archive for potential reddened targets.   
The names, coordinates, spectral types, and spectral type references for the sample are given in Table~\ref{tab_sample}.
The sources were taken from various catalogs specifically AzV \citep{Azzopardi75}, 2dFS \citep{Evans04}, MR12 \citep{MaizApellaniz12}, NGC330/NGC346 \citep{Evans06}, SMC5 \citep{Martayan07}, BBB \citep{Basinski67}, and SK \citep{Sanduleak69}.

This sample represents the full set of SMC extinction curves currently measured using UV slit spectroscopy observations.
The six sightlines in the STIS1 (PID 8198) program are described by \citet{Gordon03}.
The sightlines in the STIS2 (PID 12258) program started with an initial sample of reddened OB main sequence and giant stars from the 2dF and VLT-FLAMES surveys \citep{Evans04, Evans06, Martayan07}.
This initial sample was culled of stars with photometric mid-infrared excess (Be star, winds, etc.) using the SAGE-SMC survey \citep{Gordon11}.  
A final sample of 15 reddened stars was chosen based on the desire to have high extinctions, span the spectral type range, spatially span the galaxy, and obtain HST/STIS observations in a reasonable amount of observing time.
The planned four sightlines in the STIS3 (PID 14225) program were those in \citep{MaizApellaniz12} with significant extinction.
Unfortunately, the planned observations of one of these four (MR12~08) ended up observing the nearby multiple source MR12~06/07 that is only lightly extinguished, which was the brightest source in the STIS target acquisition 5x5\arcsec\ field-of-view.
Thus, only three of the four planned MR12 sightlines were observed.

Of the 35 stars in the sample, it was not possible to measure extinction curves for three of them (AzV~23, AzV~404, and MR12~06/07), due to their very low \ebv\ values.
Thus, while there are 35 stars in the sample, only 32 extinction curves are presented.
The eight extinction curves previously published are designated in Table~\ref{tab_sample} with table notes giving the references.
The UV spectra for all the program stars are plotted in Fig.~\ref{fig_uv_spectra}.
The details of the observational details and reduction of the IUE spectra are given by \citet{Gordon98}.  
The observational details and data reduction for the STIS spectra are given in the next section.

\subsection{STIS Observations}

\begin{deluxetable*}{lccccc}
\tabletypesize{\footnotesize}
\tablecaption{HST STIS Observations \label{tab_obs_details}}
\tablehead{ & & \multicolumn{2}{c}{FUV} & \multicolumn{2}{c}{NUV} \\
\colhead{Name} & \colhead{Aperture} & \colhead{Rootname} & \colhead{Exposure (s)} &
\colhead{Rootname} & \colhead{Exposure (s)} }
\startdata
2DFS 413   & 52X2    &objw07020 	 &1729   &objw07010          &579\\
2DFS 626   & 2X2     &objw06020 	 &1919   &objw06010          &389\\
2DFS 662   & 2X2     &objw09020 	 &1931   &objw09010          &401\\
2DFS 699   & 52X2,2X2&objw11030,objw11040&850,850&objw11010,objw11020&150,150\\
2DFS 3014   & 52X2    &objw08020          &1729   &objw08010          &579\\
2DFS 3030   & 52X2    &objw01020          &1919   &objw01010          &389\\
2DFS 3171   & 52X2    &objw04020          &1725   &objw04010          &575\\
AzV 4       & 52X0.5  &o5cg01030,o5cg01040&540,540&o5cg01010,o5cg01020&360,360\\
AzV 23      & 52X0.5  &o5cg05010,o5cg05020&540,540&o5cg05030,o5cg05040&360,360\\
AzV 86      & 52X0.5  &o5cg04010--50\tablenotemark{a} &120,100,100,120,120 &o5cg04060--a0&120(5)\\
AzV 132     & 52X0.5  &o5cg03010,o5cg03020&540,540&o5cg03030,o5cg03040&360,360\\
AzV 218     & 52X0.5  &o5cg02010,o5cg02020&540,540&o5cg02030,o5cg02040&360,360\\
AzV 404     & 52X0.5  &o5cg06010--50\tablenotemark{a} &144(5) &o5cg06060--a0      &120(5)\\
MR12 06/07  & 52X2    &ocxn01020             &1547   &ocxn01010          &700  \\ 
MR12 09     & 52X2    &ocxn02020             &1477   &ocxn02010          &780  \\
MR12 10     & 52X2    &ocxn03020             &1535   &ocxn03010          &712  \\ 
MR12 11     & 52X2    &ocxn04020             &1455   &ocxn04010          & 800\\  
NGC330 ELS 110 & 52X2    &objw14020          &1415   &objw14010          &885\\
NGC330 ELS 114 & 52X2    &objw13020          &1533   &objw13010          &767\\
NGC330 ELS 116 & 52X2    &objw15020          &1415   &objw15010          &885\\
NGC346 ELS 056 & 52X2    &objw12020          &1312   &objw12010          &1012 \\
SMC5-398& 2X2     &objw05020          &1801   &objw05010          &523\\
SMC5-3739& 52X2,2X2&objw02030,objw02040&850,850&objw02010,objw02020&150,150\\
SMC5-79264& 52X2    &objw10020          &1733   &objw10010          &397\\ 
SMC5-82923& 2X2     &objw03020          &1801   &objw03010          &523
\enddata
\tablenotetext{a}{A sequence of 5 consecutive exposures for both G140L and G230L, where 120(5) indicates five exposures of 120~s.}
\end{deluxetable*}

\begin{figure*}[tbp]
\epsscale{1.15}
\plottwo{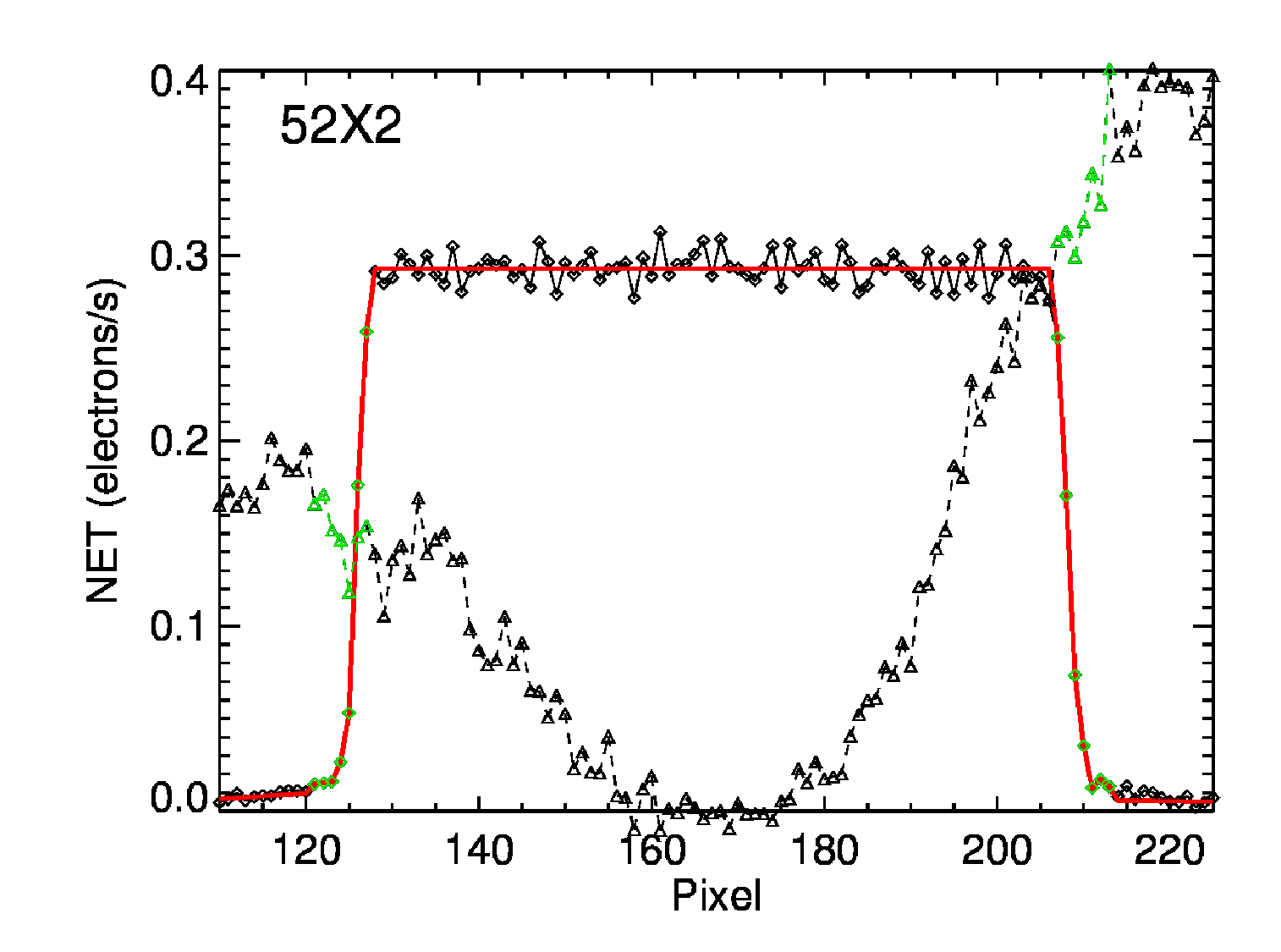}{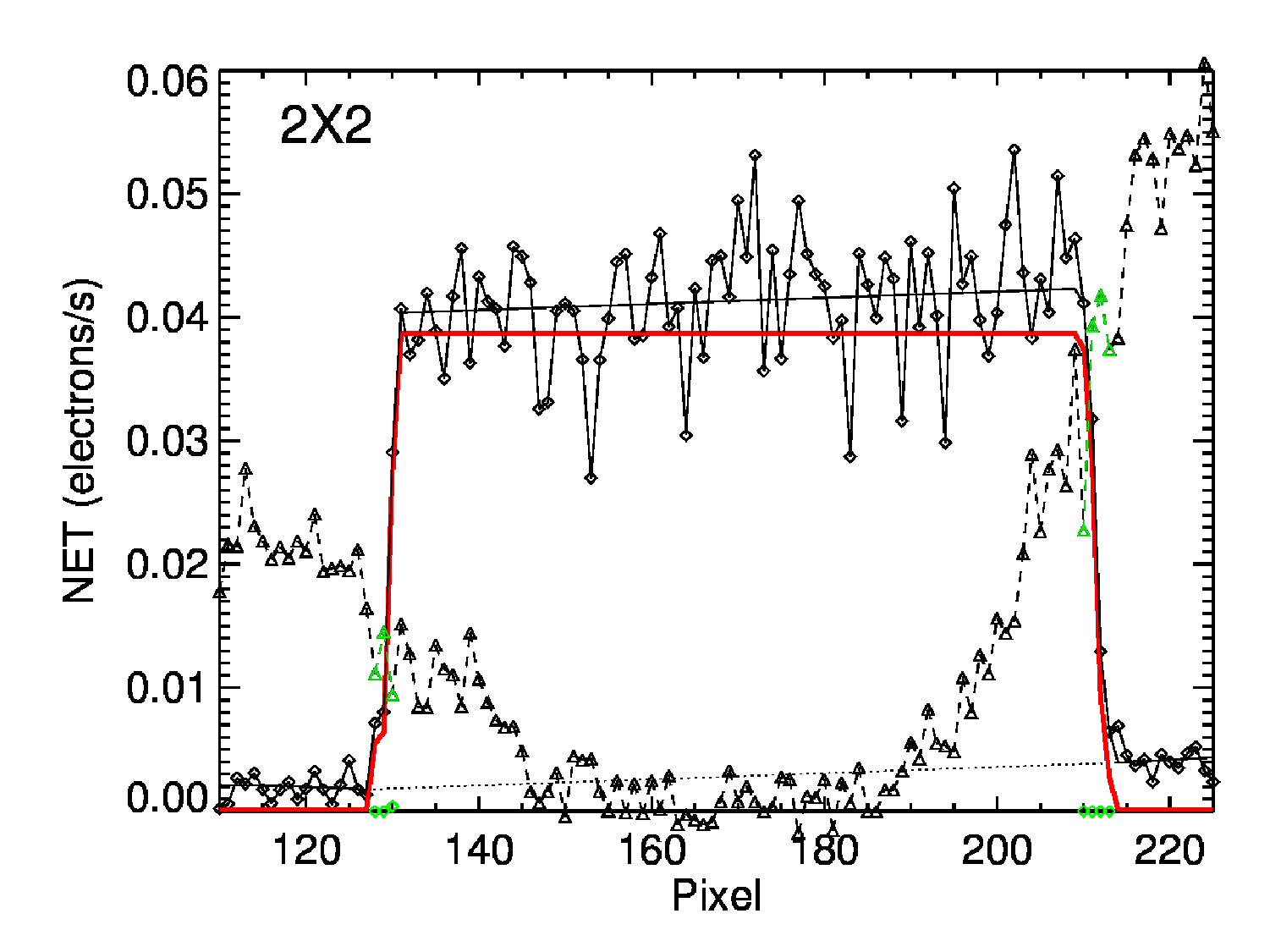}
\caption{
The sky background (diamonds) in the Ly$\alpha$ region and resulting final net signal (triangles) after subtracting the sky for an observation using the 52X2 slit (left, objw13020, NGC330 ELS 114) and one using the 2X2 slit (right, objw11040, 2DFS 0699).
For the case of the long 52x2 slit, the sky background signal (red line) in the emission line regions is the measured average at 300  pixels above and below the spectral trace, except for the steep edges of the Ly$\alpha$ slit image (green points).  
To isolate the pure Ly$\alpha$ sky signal (red line) in the case of the 2X2 slit, \kchange{the solid grey line fit to the diamonds is corrected for the stellar contribution by the dotted grey line} that is fit to the pure stellar signal outside the emission region and used as an interpolation across the broad geo-coronal Ly$\alpha$ sky region.
\label{rcb52x2}}
\end{figure*}

STIS spectra of our program stars cover 1150--3180 \AA\ at $R=500-1000$ and utilize the low resolution G140L and G230L gratings with the 52X0.5, 52X2 or 2X2 slit (all slit dimensions are in arcsec).
Our goal is to detect the 300-400~\AA\ wide 2175~\AA\ feature at 5$\sigma$ for a 5\% absorption depth.
This goal translates to a signal-to-noise of 100 at $R = 10$ and defined the exposure times.
For the brighter sources in our sample, longer exposure times were used to fill the HST orbit and potentially reveal weaker 2175~\AA\ bumps.

The older observations of the AzV stars from HST proposal 8198 utilized the 52X0.5 slit, while the newer data from HST proposal 12258 used $2\arcsec$ wide slits, which provide higher photometric accuracy \citep{Bohlin98}.
The avoidance of bright objects in the long 52X2 slit sometimes required the 2X2 entrance aperture; and the observations of 2DFS~699 and SMC5-3739 in both slits demonstrate the equivalence of the absolute fluxes.
The geocoronal emission from Ly$\alpha$ at 1215.67~\AA\ and from the much weaker OI line at 1302--1306~\AA\ are imaged on top of the stellar signal; and the total background signal must be subtracted from the extracted gross signal to get the pure net stellar signal.
The default background is the average of an upper and lower signal located 300 pixels above and below the spectral trace on the 1024x1024 pixel STIS images.

In the case of the 52X2 slit, the geocoronal sky is imaged and measured along with the residual background at all 1024 pixels.
The background signal shortward of Ly$\alpha$ is fit with a cubic polynomial to the measurements from pixel 2--120, while a quartic polynomial is fit to the measurements longward of Ly$\alpha$ with the OI region excluded, i.e., the fit is for pixels 214--272 and 365--1023.
The backgrounds in the lines at pixels 128--206 and at 280--356 are the averages in those two regions.
Because the pixel ranges of the geocoronal lines varies by several pixels and because the emission profiles do not have perfectly sharp edges, there are a few pixels where the measured background must be used without any reduction of noise, i.e. at pixel ranges 121--127, 207--213, 273--279, and 357--364.
For example, Figure~\ref{rcb52x2} shows the measured background, the fitted background, and the net=gross-background in the Ly$\alpha$ region.

In the case of the 52X0.5 slit, the geocoronal sky is flat within the noise at 159--175 and 311--326 pixel regions, while the background can be smoothly fit at 2--153, 182--272, and beyond 365.

In the case of the short 2X2 slit, the measured sky at 300 pixels from the spectral trace has no emission line signal and can be fit with one quartic polynomial.
In order to estimate the emission line strengths for adding to this standard baseline, a near background is extracted at 32 pixels ($\sim$0.8\arcsec) from the center with a width of 11 pixels, which keeps this background extraction region within the 2\arcsec\ slit.
Following the procedure used for the 52X2 slit and due to the wings of the STIS PSF extending into the near background region, for the 2x2 slit this region requires a cubic at pixels 2--127 and a quartic at 214--272 and 365--1023.
This near background as interpolated across the sky lines defines the subtraction at pixels 128--213 and at 273--364, which is added to the measured sky at 300 pixels to get the final total background that is subtracted from the gross signal.
For example, the flat part of the Ly$\alpha$ in this difference is at pixels 131--209, as illustrated by the red line in Figure~\ref{rcb52x2}.

\subsection{Ancillary Data}
\label{sec_anc_data}

The literature compilation of the optical and near-infrared photometry for the sample stars is given in Appendix~\ref{data_photometry}.

\begin{deluxetable}{lcccccc} 
\tabletypesize{\footnotesize}
\tablecaption{Ancillary Data} \label{tab_anc}
\tablehead{ \colhead{Name} & \colhead{$N_{MW}(HI)$} & \colhead{$E(B-V)_{MW}$} &
  \colhead{$q_{PAH}$} \\ & [$10^{20}$ cm$^{-2}$] & [mag] & [\%]}
\startdata
2dFS 413 & $3.224 \pm 0.079$ & $0.039 \pm 0.0010$ & $0.53 \pm 0.40$ \\ 
2dFS 626 & $3.183 \pm 0.087$ & $0.038 \pm 0.0010$ & $0.87 \pm 0.21$ \\ 
2dFS 662 & $3.179 \pm 0.060$ & $0.038 \pm 0.0007$ & $1.01 \pm 0.26$ \\ 
2dFS 699 & $3.528 \pm 0.045$ & $0.043 \pm 0.0005$ & $0.52 \pm 0.45$ \\ 
2dFS 3014 & $2.492 \pm 0.035$ & $0.030 \pm 0.0004$ & $0.32 \pm 0.14$ \\ 
2dFS 3030 & $2.367 \pm 0.069$ & $0.029 \pm 0.0008$ & $0.86 \pm 0.21$ \\ 
2dFS 3171 & $2.368 \pm 0.063$ & $0.029 \pm 0.0008$ & $1.09 \pm 0.19$ \\ 
AzV 4 & $3.499 \pm 0.035$ & $0.042 \pm 0.0004$ & $1.87 \pm 1.49$ \\ 
AzV 18 & $3.179 \pm 0.087$ & $0.038 \pm 0.0010$ & $0.57 \pm 0.19$ \\ 
AzV 23 & $3.208 \pm 0.064$ & $0.039 \pm 0.0008$ & $1.31 \pm 0.21$ \\ 
AzV 70 & $3.621 \pm 0.048$ & $0.044 \pm 0.0006$ & $1.28 \pm 0.45$ \\ 
AzV 86 & $3.156 \pm 0.030$ & $0.038 \pm 0.0004$ & $0.83 \pm 0.48$ \\ 
AzV 132 & $3.134 \pm 0.062$ & $0.038 \pm 0.0007$ & $0.71 \pm 0.22$ \\ 
AzV 214 & $3.019 \pm 0.061$ & $0.036 \pm 0.0007$ & $0.31 \pm 0.16$ \\ 
AzV 218 & $3.175 \pm 0.071$ & $0.038 \pm 0.0009$ & $0.61 \pm 0.28$ \\ 
AzV 289 & $3.021 \pm 0.035$ & $0.036 \pm 0.0004$ & $0.64 \pm 0.46$ \\ 
AzV 380 & $2.576 \pm 0.036$ & $0.031 \pm 0.0004$ & $1.19 \pm 0.66$ \\ 
AzV 398 & $2.911 \pm 0.050$ & $0.035 \pm 0.0006$ & $0.60 \pm 0.25$ \\ 
AzV 456 & $2.666 \pm 0.031$ & $0.032 \pm 0.0004$ & $0.92 \pm 0.50$ \\ 
AzV 462 & $2.766 \pm 0.029$ & $0.033 \pm 0.0003$ & $1.48 \pm 0.78$ \\ 
BBB SMC280 & $3.180 \pm 0.061$ & $0.038 \pm 0.0007$ & $1.13 \pm 0.19$ \\ 
MR12 09 & $3.185 \pm 0.032$ & $0.038 \pm 0.0004$ & $1.11 \pm 0.18$ \\ 
MR12 10 & $3.184 \pm 0.032$ & $0.038 \pm 0.0004$ & $1.11 \pm 0.18$ \\ 
MR12 11 & $3.183 \pm 0.032$ & $0.038 \pm 0.0004$ & $1.11 \pm 0.18$ \\ 
NGC330 ELS 110 & $3.263 \pm 0.044$ & $0.039 \pm 0.0005$ & $1.07 \pm 0.41$ \\ 
NGC330 ELS 114 & $3.250 \pm 0.038$ & $0.039 \pm 0.0005$ & $0.57 \pm 0.28$ \\ 
NGC330 ELS 116 & $3.273 \pm 0.046$ & $0.039 \pm 0.0006$ & $0.92 \pm 0.30$ \\ 
SK 191 & $3.437 \pm 0.046$ & $0.041 \pm 0.0006$ & \nodata \\ 
SMC5-398 & $3.343 \pm 0.046$ & $0.040 \pm 0.0006$ & $0.33 \pm 0.21$ \\ 
SMC5-3739 & $3.188 \pm 0.043$ & $0.038 \pm 0.0005$ & $0.75 \pm 0.17$ \\ 
SMC5-79264 & $3.208 \pm 0.037$ & $0.039 \pm 0.0004$ & $0.63 \pm 0.22$ \\ 
SMC5-82923 & $3.238 \pm 0.052$ & $0.039 \pm 0.0006$ & $0.57 \pm 0.42$ \\ 
\enddata
\end{deluxetable}

Because we use stellar atmosphere models to measure extinction curves (\S\ref{sec_ext_curves}), the result will be curves that are a combination of MW foreground and SMC internal dust.
Thus, we need to correct for the MW foreground dust extinction to isolate the SMC dust extinction curves.
The amount of MW foreground dust can be estimated based on radio \ion{H}{1} measurements integrated over MW velocities along each of our sightlines.
Such measurements are available with a spatial resolution of 16\arcmin\ \citep{McClure-Griffiths09, Kalberla10, Kalberla15}.
These measurements are given in Table~\ref{tab_anc} for all our sightlines where $N_{MW}(HI)$ is determined by integrating the velocity resolved spectrum over the MW velocities of -70 to 70 \kms.
The MW foreground dust column can be estimated by calculating \ebvmw\ using the $N_{MW}(HI)$ values and the MW high Galactic latitude measured $N(HI)/\ebv$ value of $8.3 \times 10^{21}$~H~cm$^{-2}$~mag$^{-1}$ \citep{Liszt14}.
Note this value is higher than that measured at low Galactic latitude $5 \times 10^{21}$~H~cm$^{-2}$~mag$^{-1}$ \citep{Bohlin78, Diplas94, Liszt14}. 

The resulting \ebvmw\ are given in Table~\ref{tab_anc}. 
The \ebvmw\ values range from 0.029 to 0.044~mag with an average of $0.037 \pm 0.0037$~mag.
This is significantly lower than the earlier estimate by \citep{Schwering91} ($\ebvmw = 0.07 - 0.09$), also based on \ion{H}{1}, but compatible with the results of \citep{Bell19} ($\ebvmw = 0.034 \pm 0.011$), based on the  \citet{Schlegel98} dust maps.
As will be seen in \S\ref{sec_ext_curves} the lower values produce more reasonable foreground corrected extinction curves (e.g., fewer nonphysical negative \fbump\ bumps).

One of the goals of this work is to compare the strength of the \fbump\ bump versus the implied dust mass fraction of PAH grains, \qpah, defined as carbonaceous grains with $< 10^3$ C atoms \citep{Draine07model}, to explain the emission seen in the MIR carbonaceous features (aka PAH features).
Fortunately, \qpah\ values for the majority of our sightlines can be measured using the same data and models as used by \citet{Chastenet19}.
A new \qpah\ map was determined by fitting the Spitzer and Herschel observations at a resolution 42\arcsec\ with a data quality criteria of $1\sigma$ (instead of $3\sigma$) in all bands.
This allowed for almost all of our sightlines to be associated with a \qpah\ measurements.
The \qpah\ values and uncertainties for our sightlines are given in Table~\ref{tab_anc}.

\section{Extinction Curves}
\label{sec_ext_curves}

In this paper, we utilize the ``Extinction-Without-Standards'' technique \citep{Fitzpatrick05, Fitzpatrick19} to model the observed spectral energy distributions (SEDs) of our sample of reddened stars and, ultimately, produce extinction curves covering the UV through the near-IR spectral region.
In contrast to the ``Pair Method'' \citep[e.g.,][]{Massa83}, which requires observations of unreddened standard stars, this technique uses stellar atmosphere calculations to represent the intrinsic SEDs of the reddened stars.
This eliminates the need to observe a large grid of unreddened stars and allows the effects of mismatch between the stellar properties (e.g., \teff\ or \logg) and the extinction curves to be investigated.
On the other hand, it places heavy reliance on the accuracy of the atmosphere models and the absolute calibration of the observations.
In addition, the Extinction-Without-Standards curves include contributions from all the dust in their respective sightlines - in this case, the Milky Way foreground - whereas such ``contamination'' would be largely removed by the Pair Method.
This drawback is outweighed by elimination of standard star observations and also offers the opportunity to investigate the extinction properties of the foreground dust.

\subsection{SED Modeling}
\label{sec_stellar_model}

\begin{deluxetable*}{lccccc}
\tablecaption{Stellar Fit Parameters \label{tab_stellar_param}}
\tablehead{\colhead{Name} & \colhead{$\ebv_\mathrm{MW+SMC}$} &\colhead{\teff\ [K]} & \colhead{\logg} & \colhead{Model Grid\tablenotemark{a}}}
\startdata
2dFS 413  & $0.200 \pm 0.025$ & $19172 \pm 1060$ & $3.186 \pm 0.406$ & TLUB2 \\
2dFS 626  & $0.247 \pm 0.021$ & $23518 \pm 998$ & $4.3$\tablenotemark{b}  & TLUB2 \\
2dFS 662  & $0.237 \pm 0.018$ & $26919 \pm 416$ & $4.3$\tablenotemark{b} & TLUB2 \\
2dFS 699  & $0.184 \pm 0.020$ & $15216 \pm 554$ & $2.232 \pm 0.255$ & TLUB2 \\
2dFS 3014 & $0.236 \pm 0.024$ & $25953 \pm 765$ & $4.3$\tablenotemark{b} & TLUB2 \\
2dFS 3030 & $0.287 \pm 0.024$ & $36226 \pm 301$ & $4.3$\tablenotemark{b} & TLUO10\\
2dFS 3171 & $0.400 \pm 0.019$ & $35412 \pm 929$ & $4.3$\tablenotemark{b} & TLUO10 \\
AzV 4     & $0.315 \pm 0.006$ & $24156 \pm 249$ & $2.766 \pm 0.049$ & TLUB10 \\
AzV 18    & $0.202 \pm 0.011$ & $19636 \pm 326$ & $2.335 \pm 0.081$ & TLUB10 \\
AzV 23    & $0.249 \pm 0.008$ & $18076 \pm 395$ & $3.090 \pm 0.137$ & TLUB10\\
AzV 70    & $0.099 \pm 0.015$ & $30414 \pm 547$ & $3.379 \pm 0.099$ & TLUO10 \\
AzV 86    & $0.122 \pm 0.014$ & $28774 \pm 148$ & $4.3$\tablenotemark{b} & TLUO10 \\
AzV 132   & $0.385 \pm 0.016$ & $33343 \pm 172$ & $4.3$\tablenotemark{b} & TLUO10 \\
AzV 214   & $0.232 \pm 0.014$ & $22452 \pm 775$ & $2.660 \pm 0.167$ & TLUB10 \\
AzV 218   & $0.163 \pm 0.017$ & $24212 \pm 300$ & $2.982 \pm 0.077$ & TLUB10 \\
AzV 289   & $0.114 \pm 0.016$ & $28193 \pm 726$ & $3.111 \pm 0.127$ & TLUO10 \\
AzV 380   & $0.090 \pm 0.016$ & $21834 \pm 544$ & $2.554 \pm 0.127$ & TLUB10 \\
AzV 398   & $0.322 \pm 0.021$ & $30229 \pm 1329$ & $3.031 \pm 0.156$ & TLUO10 \\
AzV 456   & $0.387 \pm 0.011$ & $34525 \pm 845$ & $4.3$\tablenotemark{b} & TLUO10 \\
AzV 462   & $0.051 \pm 0.022$ & $20705 \pm 384$ & $2.497 \pm 0.140$ & TLUB10 \\
BBB SMC280 & $0.193 \pm 0.018$ & $22773 \pm 1843$ & $2.990 \pm 0.705$ & TLUB2 \\
MR12 09  & $0.133 \pm 0.005$ & $15886 \pm 300$ & $4.3$\tablenotemark{b} & TLUB2 \\
MR12 10  & $0.249 \pm 0.010$ & $14465 \pm 500$ & $4.3$\tablenotemark{b} & BOSZ \\
MR12 11  & $0.309 \pm 0.008$ & $20410 \pm 400$ & $4.3$\tablenotemark{b} & TLUB2 \\
NGC330 ELS 110 & $0.093 \pm 0.007$ & $26247 \pm 520$ & $4.3$\tablenotemark{b} & TLUB2 \\
NGC330 ELS 114 & $0.097 \pm 0.012$ & $24051 \pm 599$ & $4.3$\tablenotemark{b} & TLUB2 \\
NGC330 ELS 116 & $0.080 \pm 0.009$ & $20899 \pm 495$ & $4.3$\tablenotemark{b} & TLUB2 \\
SK 191     & $0.119 \pm 0.019$ & $19863 \pm 307$ & $2.270 \pm 0.095$ & TLUB10 \\
SMC5-398   & $0.147 \pm 0.011$ & $16790 \pm 279$ & $4.3$\tablenotemark{b} & TLUB2 \\
SMC5-3739  & $0.141 \pm 0.011$ & $15964 \pm 411$ & $3.474 \pm 0.241$ & TLUB2 \\
SMC5-79264 & $0.231 \pm 0.017$ & $17505 \pm 470$ & $4.220 \pm 0.361$ & TLUB2\\
SMC5-82923 & $0.129 \pm 0.020$ & $16410 \pm 372$ & $3.714 \pm 0.216$ & TLUB2
\enddata
\tablenotetext{a}{This column indicates the model atmosphere grid used to compute the intrinsic SED each star.
``TLUO10'' refers to the nLTE O star grid of \citet{Lanz03}, computed with a value of \vturb\ = 10 \kms.
``TLUB2'' and "TLUB10'' refer to the B star grids from \citet{Lanz07}, computed with \vturb = 2 \kms\ and 10 \kms, respectively.
In the \teff\ and \logg\ region where these models overlap, the model which provided the best fit to the data was adopted.
``BOSZ'' refers to LTE grid of \citet{Bohlin17}, computed with \vturb\ = 2 \kms.
These models were used for the one star with \teff $\le$ 15000 K.}
\tablenotetext{b}{\logg\ fixed due to insufficient spectral information.}
\end{deluxetable*}

As described in detail by \citet{Fitzpatrick05, Fitzpatrick19}, Extinction-Without-Standards is a modeling process in which the observed UV through near-IR SEDs of the reddened stars are fit with a combination of a stellar atmosphere model, interstellar \ion{H}{1} Ly$\alpha$ absorption and a parameterized representation of the wavelength dependence of the total sightline extinction.
As in \citet{Fitzpatrick19}, the principal goal of the modeling is to utilize the full range of available spectral information to determine the stellar properties and thus the intrinsic SED.
The actual extinction curves are derived from normalized ratio of the observed and intrinsic SEDs, analogously to the Pair Method.   

In this study, we utilized the O and B star TLUSTY \kchange{non-Local Thermodynamic Equilibrium (nLTE)} atmosphere grids \citep{Lanz03, Lanz07} for all of our targets except MR12~10, whose \teff\ is below the lower limit of the \kchange{nLTE} TLUSTY grid \kchange{of 15000~K}.
For this star, we used the BOSZ LTE models \citep{Bohlin17}.
\kchange{TLUSTY models are strongly preferred for stars with $\teff \geq 15000~K$ as accounting for nLTE effects are important for accurate predictions of their spectra.}
The interstellar Lyman~$\alpha$ profile was taken from \citet{Bohlin75}.
The wavelength dependence of the extinction was represented in the optical and near-IR ($\lambda > 4000~\mathrm{\AA}$) using the extinction curve of \citet{Fitzpatrick19} and in the UV ($\lambda < 2700~\mathrm{\AA}$) with the fitting function of \citet{Fitzpatrick07}.
These two pieces of the overall curve were joined together by a cubic spline interpolation.
The determination of the best-fit model proceeded as follows: initial estimates of the various parameters were used to produce a high resolution model of the reddened star's SED.
In the UV, the model was then convolved with the appropriate instrumental line spread function (i.e., STIS or IUE) and binned to match the observed data.
In the optical and near-IR, synthetic photometry was performed on the bandpasses available for the star (see Appendix~\ref{data_photometry}).
A value of $\chi^2$ was computed, characterizing the goodness of the fit.
The various parameters were then iteratively adjusted to arrive at the best possible fit to the data, as indicated by the $\chi^2$ value.
The iteration and minimization were managed by the gradient-search algorithm MPFIT written in the Interactive Data Language (IDL) by C. Markwardt\footnote{\url{https://pages.physics.wisc.edu/~craigm/idl/fitting.html}}.

The typical SED fits required the determination of 15 free parameters.
These included three for the stellar model (\teff, \logg, and \vrad), one for Ly$\alpha$ profile of SMC gas (\nhismc), and 10 for the extinction curve. 
The extinction curve parameters are $E(44-55$) and $R(55)$ to specify the optical/NIR portion, the seven FM07 parameters to specify the UV portion, and a floating spline anchor point at 3000 \AA\ to facilitate the joining of the UV and optical.
Note that $E(55-44) \sim \ebv$ and $R(55) \sim \rv$ \citep{Fitzpatrick19} and we use the more common \ebv\ and \rv\ throughout this paper.
In addition, a single scale factor is determined, to align the models with the observations at a wavelength of 5500 \AA.
In addition to these 15 free parameters, there are number of fixed parameters.
Due to the relatively low spectral resolution of the spectrophotometry, we assumed a stellar metallicity of $\abund = -0.7$ and a rotational velocity of $\vrot = 100~\kms$ for all the fits.
The values of the microturbulence (either 2~\kms\ or 10~\kms) were determined by the appropriate model grid.
We also assumed that the MW foreground interstellar \ion{H}{1} component is located at 0~\kms\ with N$_\mathrm{MW}$(HI) = $3.5\times 10^{20}$~H~cm$^{-2}$ and that the SMC component is located at 120~\kms.
Finally, the line-spread functions for the spectrophotometric observations were taken as Gaussians with FWHM values of 1.2 \AA\ for STIS/G140M, 3.2 \AA\ for STIS/G230M, and 5.5 \AA\ for IUE.

The best fit values for the total reddening toward each star (i.e, MW plus SMC) and the stellar parameters \teff\ and \logg\ are given in Table~\ref{tab_stellar_param}, along with the specific model grid used.
In some cases, there was insufficient information to determine definitive values of \logg.
This usually occurs for the higher gravity stars, in which the Balmer continuum -- the main gravity indicator in these fits -- is less sensitive to \logg.
In these cases, we fixed the gravity to a value of $\logg = 4.3$.
Although we also fit for values of \vrad\ for each star, the low spectral resolution coupled with uncertainties in the wavelength calibration zero points render these values of little interest.
The parameters describing the shape of the total UV through near-IR extinction curves are discarded at this point, as their sole purpose was to facilitate the SED modeling.
The actual extinction curves analyzed below are created by ratioing the observed and best-fit intrinsic SEDs and normalizing by the appropriate value of \ebv.

\subsection{MW Foreground Correction}
\label{sec_foreground}

\begin{figure*}[tbp]
\epsscale{1.15}
\plottwo{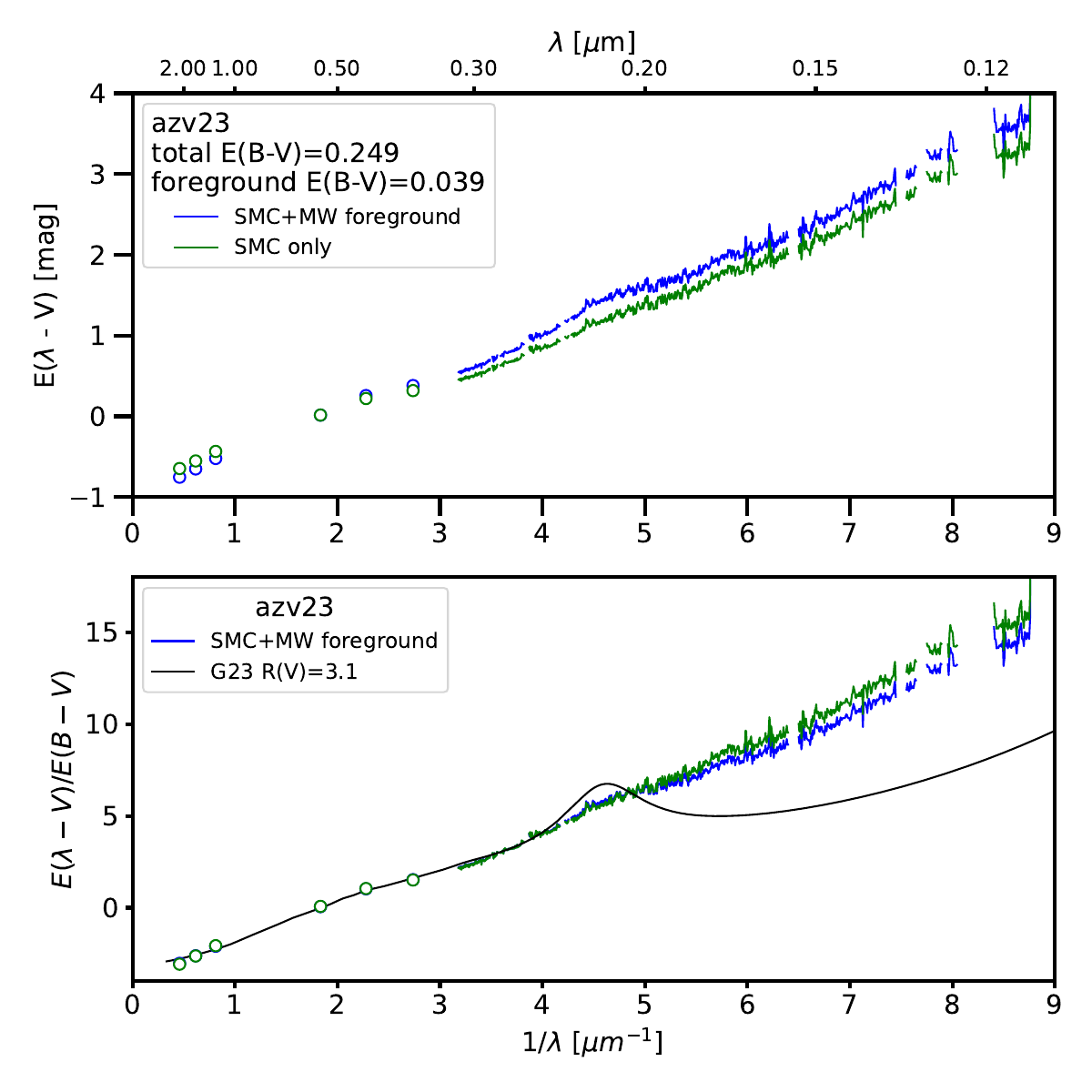}{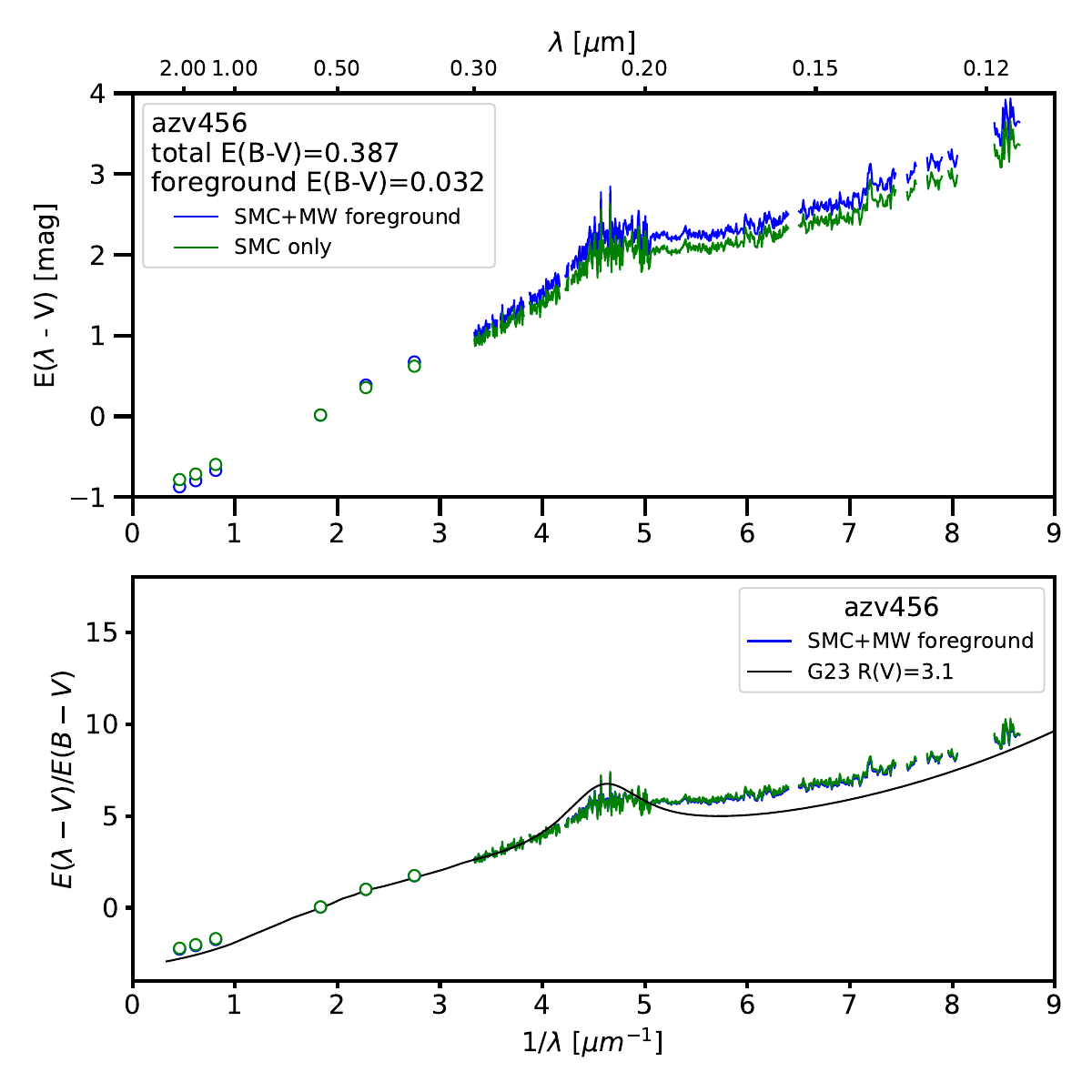}
\caption{The as measured (SMC+MW) and after MW foreground corrected (SMC only) extinction curves are shown and illustrate the importance of the foreground correction.
The upper plots give the as measured extinction curves and the bottom plots the normalized extinction curves.
The foreground correction is small, but significant especially for AzV 23 where the very weak \fbump\ bump is seen to be due to MW foreground dust.
\kchange{The curves are plotted versus $1/\lambda$ as this is common for UV extinction studies and it is particularly useful as the majority of the SMC sightlines are roughly linear with $1/\lambda$}.
\label{fig_forecor_example}}
\end{figure*}

Extinction curves produced from the ratio of the intrinsic SEDs derived above and the observed SMC SEDs contain contributions from the SMC internal dust and MW foreground dust.
For this paper we focus on the properties of SMC dust and so need to correct for the MW foreground extinction.
This correction was done by subtracting a model of the MW foreground extinction from the measured total extinction curve.
The model used was the MW average curve corresponding to the Galactic average value of $R(V) = 3.1$ \citep{Gordon23} and the appropriate MW foreground reddening (\ebvmw; see Table \ref{tab_anc}).
To the extent that the wavelength dependence of the foreground extinction can represented by the $R(V) = 3.1$ curve, the result will be SMC-only extinction curves.
The normalization values for the resulting SMC only curves, \ebvsmc, are determined by subtracting the \ebvmw\ values in Table~\ref{tab_anc} from the total $\ebv_\mathrm{MW+SMC}$ values in Table~\ref{tab_stellar_param}.
While little is known of the true wavelength dependence of MW extinction extinction along high latitude sightlines, the success of this model indicates that it is likely similar to MW average.
In future work, we plan to investigate the MW foreground extinction along Magellanic Cloud sightlines using observations of the most lightly reddened stars in the SMC and LMC.

The effect of the MW foreground correction is illustrated in Fig.~\ref{fig_forecor_example} for two sightlines.
The first sightline, towards AzV~23, shows a weak \fbump\ in the measured MW+SMC extinction curve, which is eliminated once the MW foreground is removed.
This behavior is seen for many of our sightlines, i.e., weak \fbump\ signatures that disappear once the MW foreground is removed.
This suggests the our assumed form for the MW foreground extinction curve does not overestimate the normalized strength of the \fbump\ bump, otherwise the corrected curves would show dips at \fbump.
The behavior for the AzV~23 curve is in contrast to the sightline towards AzV~456 where the MW foreground correction changes the normalized extinction curve very little.
This is not surprising as the assumed MW foreground extinction curve is similar to the combined MW+SMC extinction curve for this sightline.

\subsection{The SMC Extinction Curves}
\label{sec_final_curves}

\begin{figure*}[tbp]
\epsscale{1.15}
\plotone{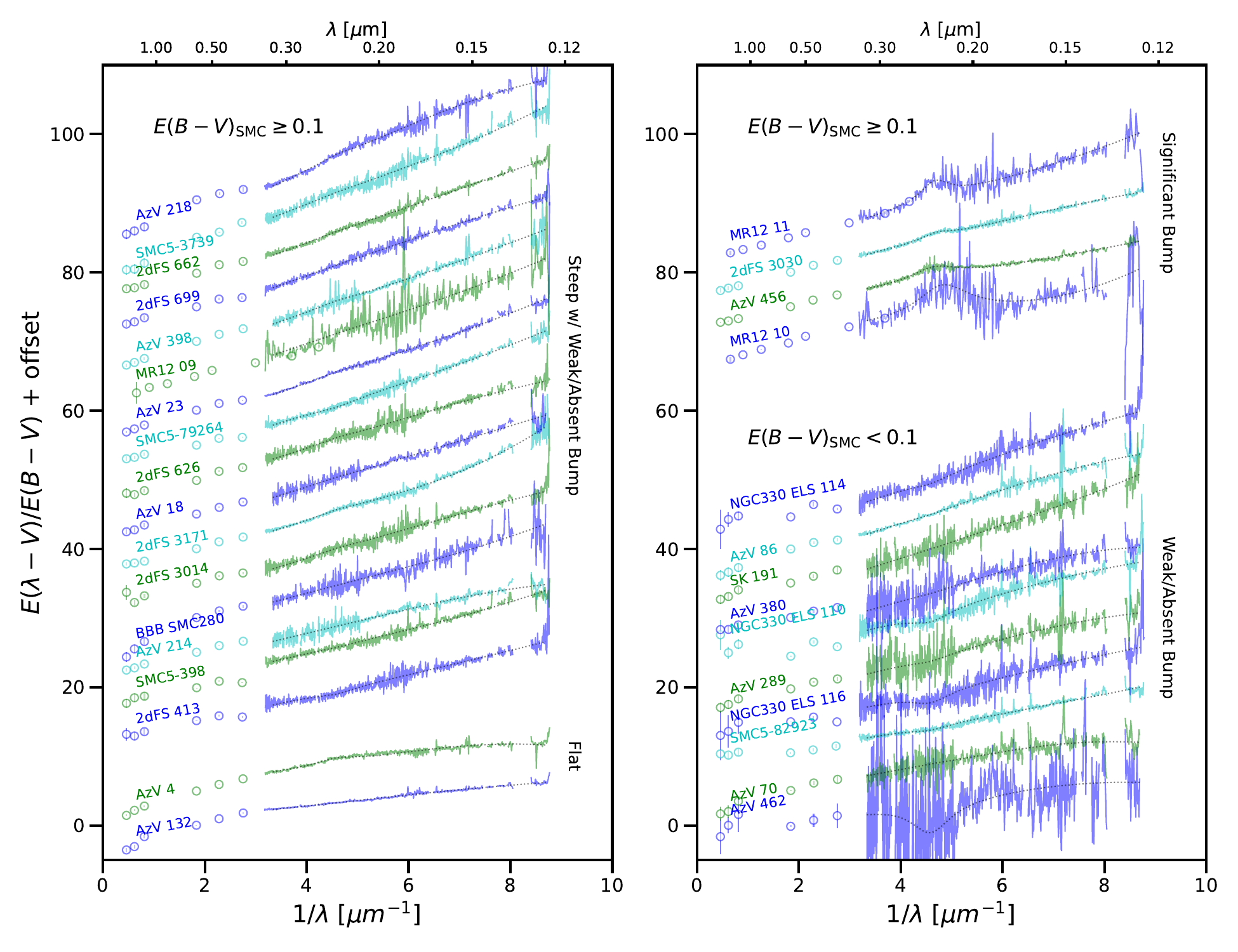}
\caption{The MW foreground corrected SMC only extinction curves are plotted along with the FM90 fits for the UV portion (dotted lines).
For clarity, the curves have been offset on the y-axis by multiples of 5.
For ease for interpretation the top axis labels provide representative values in $\lambda$ units.
Regions of strong residuals due to stellar or interstellar gas lines have been masked.
The extinction curves are grouped based on their observed behavior and \ebvsmc\ values.
Inside of each group, the curves are displayed from flattest to steepest based on the fitted FM90 $C_2$ values.
\label{fig_good_curves}}
\end{figure*}

\begin{deluxetable*}{lcccccccc}
\tabletypesize{\scriptsize}
\tablecaption{Extinction Parameters \label{tab_ext_col_param}}
\tablehead{\colhead{Name} & \colhead{\ebvsmc} & \colhead{\avsmc} & \colhead{\rvsmc} & \colhead{\nhismc} & \colhead{$C_1$} & \colhead{$C_2$} & \colhead{$B_3$} & \colhead{$C_4$} \\
 & \colhead{[mag]} & \colhead{[mag]} & & \colhead{[$10^{21}$ H cm$^{-2}$]} & & & }
\startdata
\multicolumn{9}{c}{$E(B-V)_\mathrm{SMC} \geq 0.1$, Steep with Weak/Absent Bump} \\ \hline
2dFS 413 & $0.16 \pm 0.03$ & $0.35 \pm 0.07$ & $2.17 \pm 0.54$ & $8.17 \pm 0.72$ & $-2.78 \pm 0.15$ & $1.61 \pm 0.05$ & $-0.61 \pm 0.24$ & $0.08 \pm 0.03$ \\
2dFS 626 & $0.21 \pm 0.03$ & $0.49 \pm 0.06$ & $2.35 \pm 0.37$ & $16.63 \pm 1.01$ & $-4.30 \pm 0.21$ & $2.22 \pm 0.07$ & $0.18 \pm 0.17$ & $-0.13 \pm 0.04$ \\
2dFS 662 & $0.20 \pm 0.02$ & $0.51 \pm 0.05$ & $2.56 \pm 0.34$ & $13.46 \pm 0.40$ & $-6.03 \pm 0.21$ & $2.59 \pm 0.06$ & $0.04 \pm 0.13$ & $-0.02 \pm 0.02$ \\
2dFS 699 & $0.14 \pm 0.02$ & $0.36 \pm 0.03$ & $2.52 \pm 0.42$ & $13.09 \pm 0.47$ & $-5.73 \pm 0.20$ & $2.56 \pm 0.07$ & $0.07 \pm 0.13$ & $-0.11 \pm 0.03$ \\
2dFS 3014 & $0.21 \pm 0.02$ & $0.55 \pm 0.05$ & $2.68 \pm 0.37$ & $12.78 \pm 0.42$ & $-4.96 \pm 0.13$ & $2.15 \pm 0.03$ & $0.15 \pm 0.09$ & $-0.08 \pm 0.03$ \\
2dFS 3171 & $0.37 \pm 0.02$ & $0.90 \pm 0.04$ & $2.42 \pm 0.16$ & $9.55 \pm 1.18$ & $-4.62 \pm 0.10$ & $2.18 \pm 0.03$ & $0.27 \pm 0.07$ & $0.62 \pm 0.02$ \\
AzV 18 & $0.16 \pm 0.02$ & $0.42 \pm 0.06$ & $2.58 \pm 0.37$ & $10.06 \pm 1.13$ & $-4.87 \pm 0.31$ & $2.21 \pm 0.09$ & $0.13 \pm 0.22$ & $0.01 \pm 0.04$ \\
AzV 23 & $0.21 \pm 0.01$ & $0.67 \pm 0.03$ & $3.17 \pm 0.21$ & $8.90 \pm 0.48$ & $-5.69 \pm 0.19$ & $2.41 \pm 0.06$ & $0.25 \pm 0.12$ & $0.16 \pm 0.02$ \\
AzV 214 & $0.20 \pm 0.02$ & $0.51 \pm 0.03$ & $2.59 \pm 0.25$ & $5.52 \pm 0.61$ & $-4.19 \pm 0.20$ & $1.76 \pm 0.04$ & $-0.05 \pm 0.16$ & $-0.23 \pm 0.03$ \\
AzV 218 & $0.12 \pm 0.02$ & $0.60 \pm 0.04$ & $4.83 \pm 0.80$ & $4.67 \pm 0.18$ & $-8.36 \pm 0.54$ & $3.27 \pm 0.17$ & $0.43 \pm 0.21$ & $-0.41 \pm 0.06$ \\
AzV 398 & $0.29 \pm 0.02$ & $1.03 \pm 0.03$ & $3.59 \pm 0.21$ & $11.13 \pm 2.39$ & $-5.54 \pm 0.16$ & $2.44 \pm 0.04$ & $0.19 \pm 0.11$ & $0.09 \pm 0.03$ \\
BBB SMC280 & $0.15 \pm 0.02$ & $0.87 \pm 0.04$ & $5.61 \pm 0.74$ & $7.62 \pm 1.83$ & $-4.10 \pm 0.24$ & $1.92 \pm 0.06$ & $0.02 \pm 0.19$ & $0.18 \pm 0.06$ \\
MR12 09 & $0.09 \pm 0.01$ & $0.28 \pm 0.18$ & $2.92 \pm 1.90$ & $8.31 \pm 1.22$ & $-4.95 \pm 0.53$ & $2.41 \pm 0.11$ & $0.02 \pm 0.38$ & $0.19 \pm 0.14$ \\
SMC5-398 & $0.11 \pm 0.01$ & $0.22 \pm 0.05$ & $2.08 \pm 0.49$ & $0.48 \pm 0.11$ & $-1.99 \pm 0.10$ & $1.73 \pm 0.05$ & $-0.18 \pm 0.19$ & $0.18 \pm 0.03$ \\
SMC5-3739 & $0.10 \pm 0.01$ & $0.53 \pm 0.05$ & $5.20 \pm 0.66$ & $4.50 \pm 0.21$ & $-6.32 \pm 0.31$ & $2.78 \pm 0.10$ & $0.16 \pm 0.17$ & $0.22 \pm 0.04$ \\
SMC5-79264 & $0.19 \pm 0.02$ & $0.39 \pm 0.03$ & $2.00 \pm 0.24$ & $11.20 \pm 0.55$ & $-4.82 \pm 0.11$ & $2.35 \pm 0.04$ & $-0.51 \pm 0.10$ & $0.17 \pm 0.02$ \\
SMC Average & \nodata & \nodata & $3.02 \pm 0.18$ & \nodata & $-5.07 \pm 0.05$ & $2.30 \pm 0.01$ & $0.12 \pm 0.03$ & $0.07 \pm 0.02$ \\
\hline \multicolumn{9}{c}{$E(B-V)_\mathrm{SMC} \geq 0.1$, Significant Bump} \\ \hline
2dFS 3030 & $0.26 \pm 0.03$ & $0.70 \pm 0.04$ & $2.72 \pm 0.30$ & $11.84 \pm 0.27$ & $-3.38 \pm 0.10$ & $1.79 \pm 0.03$ & $0.79 \pm 0.08$ & $-0.11 \pm 0.02$ \\
AzV 456 & $0.35 \pm 0.01$ & $0.85 \pm 0.02$ & $2.39 \pm 0.10$ & $1.31 \pm 0.22$ & $-1.76 \pm 0.09$ & $1.23 \pm 0.01$ & $1.81 \pm 0.04$ & $0.11 \pm 0.01$ \\
MR12 10 & $0.21 \pm 0.01$ & $0.65 \pm 0.09$ & $3.08 \pm 0.45$ & $36.17 \pm 16.54$ & $-0.02 \pm 0.56$ & $0.79 \pm 0.10$ & $4.52 \pm 0.39$ & $0.63 \pm 0.15$ \\
MR12 11 & $0.27 \pm 0.01$ & $0.71 \pm 0.10$ & $2.61 \pm 0.37$ & $10.10 \pm 2.59$ & $-4.09 \pm 0.26$ & $2.07 \pm 0.05$ & $2.68 \pm 0.19$ & $0.23 \pm 0.08$ \\
SMC Bumps & \nodata & \nodata & $2.55 \pm 0.10$ & \nodata & $-2.85 \pm 0.21$ & $1.51 \pm 0.04$ & $2.64 \pm 0.12$ & $0.25 \pm 0.06$ \\
\hline \multicolumn{9}{c}{$E(B-V)_\mathrm{SMC} \geq 0.1$, Flat} \\ \hline
AzV 4 & $0.27 \pm 0.01$ & $0.95 \pm 0.03$ & $3.47 \pm 0.13$ & $0.58 \pm 0.08$ & $-1.01 \pm 0.04$ & $1.11 \pm 0.01$ & $0.74 \pm 0.05$ & $-0.37 \pm 0.02$ \\
AzV 132 & $0.35 \pm 0.02$ & $1.17 \pm 0.07$ & $3.38 \pm 0.25$ & $6.60 \pm 0.18$ & $-0.21 \pm 0.02$ & $0.77 \pm 0.00$ & $-0.05 \pm 0.07$ & $-0.04 \pm 0.01$ \\
\hline \multicolumn{9}{c}{$E(B-V)_\mathrm{SMC} < 0.1$, Weak/Absent Bump} \\ \hline
AzV 70 & $0.06 \pm 0.02$ & $0.18 \pm 0.03$ & $3.20 \pm 1.00$ & $0.90 \pm 0.20$ & $-2.50 \pm 0.31$ & $1.35 \pm 0.07$ & $0.25 \pm 0.37$ & $-0.42 \pm 0.10$ \\
AzV 86 & $0.08 \pm 0.01$ & $0.34 \pm 0.03$ & $4.01 \pm 0.75$ & $0.06 \pm 0.05$ & $-5.89 \pm 0.23$ & $2.42 \pm 0.07$ & $-0.26 \pm 0.16$ & $-0.28 \pm 0.04$ \\
AzV 289 & $0.08 \pm 0.02$ & $0.22 \pm 0.03$ & $2.86 \pm 0.69$ & $3.20 \pm 0.23$ & $-3.42 \pm 0.41$ & $1.75 \pm 0.08$ & $-0.79 \pm 0.33$ & $-0.16 \pm 0.06$ \\
AzV 380 & $0.06 \pm 0.02$ & $0.11 \pm 0.02$ & $1.83 \pm 0.60$ & $3.37 \pm 0.62$ & $-5.03 \pm 0.50$ & $1.95 \pm 0.10$ & $-0.24 \pm 0.37$ & $-0.28 \pm 0.07$ \\
AzV 462 & $0.02 \pm 0.02$ & $0.00 \pm 0.02$ & $0.06 \pm 1.28$ & $0.07 \pm 0.04$ & $0.49 \pm 1.43$ & $0.81 \pm 0.21$ & $-4.73 \pm 1.88$ & $-0.22 \pm 0.21$ \\
NGC330 ELS 110 & $0.05 \pm 0.01$ & $-0.05 \pm 0.04$ & $-1.00 \pm 0.86$ & $3.76 \pm 0.15$ & $-3.09 \pm 0.28$ & $1.97 \pm 0.11$ & $-1.52 \pm 0.49$ & $-0.16 \pm 0.07$ \\
NGC330 ELS 114 & $0.06 \pm 0.01$ & $0.03 \pm 0.04$ & $0.57 \pm 0.76$ & $5.51 \pm 0.21$ & $-6.24 \pm 0.47$ & $2.50 \pm 0.13$ & $-0.84 \pm 0.35$ & $-0.11 \pm 0.06$ \\
NGC330 ELS 116 & $0.04 \pm 0.01$ & $0.04 \pm 0.07$ & $0.92 \pm 1.68$ & $1.98 \pm 0.13$ & $-2.85 \pm 0.36$ & $1.58 \pm 0.10$ & $-1.71 \pm 0.72$ & $-0.02 \pm 0.07$ \\
SK 191 & $0.08 \pm 0.02$ & $0.16 \pm 0.03$ & $2.12 \pm 0.63$ & $3.08 \pm 0.40$ & $-5.62 \pm 0.44$ & $2.35 \pm 0.12$ & $-0.25 \pm 0.31$ & $0.18 \pm 0.05$ \\
SMC5-82923 & $0.09 \pm 0.02$ & $-0.04 \pm 0.04$ & $-0.49 \pm 0.47$ & $3.90 \pm 0.20$ & $-1.56 \pm 0.08$ & $1.30 \pm 0.03$ & $-0.64 \pm 0.28$ & $0.05 \pm 0.03$ \\
\enddata
\end{deluxetable*}

\begin{deluxetable}{lcc}
\tabletypesize{\small}
\tablecaption{Detailed Bump Parameters\tablenotemark{a} \label{tab_ext_bump_params}}
\tablehead{\colhead{Name} & \colhead{$x_o$} & \colhead{$\gamma$} \\
 & \colhead{[\micron$^{-1}$]} & \colhead{[\micron$^{-1}$]} }
\startdata
2dFS 3030 & $4.71 \pm 0.02$ & $0.72 \pm 0.05$ \\
AzV 456 & $4.69 \pm 0.01$ & $1.42 \pm 0.07$ \\
MR12 10 & $4.83 \pm 0.04$ & $1.20 \pm 0.22$ \\
MR12 11 & $4.68 \pm 0.03$ & $0.73 \pm 0.10$ \\
SMC Bumps & $4.73 \pm 0.03$ & $1.15 \pm 0.14$ \\
\enddata
\tablenotetext{a}{For all other sightlines, these parameters were fixed to the average MW values of $x_o = 4.59$~\micron$^{-1}$ and $\gamma = 0.95$~\micron$^{-1}$.}
\end{deluxetable}

The MW foreground-corrected extinction curves for the sample of 32 stars are shown in Fig.~\ref{fig_good_curves}. 
The curves have been grouped based on their \ebvsmc\ values and shapes.
The groups are:
\begin{enumerate}
\item $\ebvsmc > 0.1$, steep UV extinction rising linearly to shorter wavelengths, and weak/absent \fbump\ bumps ($B_3 < 3\sigma$ or negative),
\item $\ebvsmc > 0.1$ and significant \fbump\ bumps ($B_3 > 3\sigma$)
\item $\ebvsmc > 0.1$, fairly flat UV extinction, and weak/absent \fbump\ bumps
\item $\ebvsmc < 0.1$ with dust columns are low enough to be significantly impacted by uncertainties in the MW foreground correction and the measured \ebvsmc.
\end{enumerate}
The first group shows behaviors similar to those labeled ``SMC Bar'' by \citet{Gordon03}.
The second group includes all the known sightlines with a 2175~\AA\ bump including AzV~456.
The third group includes two sightlines that display fairly flat UV extinction and this is a behavior that is usually associated with sightlines lacking many small grains \citep{Whittet04}.
Such a behavior could also be the result of over-estimating the \ebvsmc\ dust column or a significant mismatch between the observed and model stellar parameters.
We consider these two sightlines to be intriguing, but more work would be needed to confirm them.
The fourth group curves are generally noisier and have larger uncertainties on their overall slopes (e.g., $C_2$ values) and some display negative \fbump\ bumps with this latter property being highly unlikely given our understanding of this feature \citep[e.g.,][]{Calzetti95}.
The fourth group is included for completeness, but will not be included in any of the analysis done in this paper.

The gas columns, dust columns, and FM90 parameters for the UV portions of the extinction curves are given in Tables~\ref{tab_ext_col_param} and \ref{tab_ext_bump_params}.
In addition, average values are provided for the first and second groups.
The dust and gas columns for each extinction curve are quantified by \ebvsmc, \avsmc, and \nhismc.
The \ebvsmc\ were determined by subtracting the MW foreground reddenings in Table \ref{tab_anc} from the total reddenings in Table \ref{tab_stellar_param} that resulted from the fitting procedure.
The \avsmc\ values were determined by comparing the NIR $E(\lambda-V)_\mathrm{SMC}$ extinction to the \alav\ average from \citet{Decleir22} following the \citet{Gordon03} method updated for $A(V)$ instead of \rv, as motivated by \citet{Gordon09FUSE}.
Similar $A(V)$ values are found using the \citet{Rieke89} average as the comparison instead.
The \rvsmc\ values, which probe the average dust grain size, are $\avsmc / \ebvsmc$.
The \nhismc\ were determined from the fitting procedure with a small additional correction.
The modeling process assumed a uniform MW foreground component of $3.5\times 10^{20}~\mathrm{H}~\mathrm{cm}^{-2}$ for all stars and attributed any excess absorption in the Lyman~$\alpha$ line to SMC gas.
We corrected this result for the small difference between the assumed MW foreground column and that inferred from the 21-cm emission (Table \ref{tab_anc}).

To quantify the properties of the UV portion of each foreground-corrected curve, we fit them with the FM90 parameters, but with the \fbump\ bump strength given by $B_3 \equiv C_3 / \gamma^2$ instead of $C_3$ as in FM90.
$B_3$ is a direct measure of the \fbump\ bump amplitude and is much less correlated with $\gamma$ than $C_3$.
Given the weakness/absence of the \fbump\ bump in most of the sightlines, the \fbump\ bump center and width were fixed for all sightlines to the MW average values of 4.59~\micron$^{-1}$ and 0.95~\micron, respectively \citep{Valencic04, Fitzpatrick07, Gordon09FUSE}, except for those with significant bumps.
The FM90 fit uncertainties were determined by calculating the best fit, empirically determining the uncertainties from the differences between the observations and the best fit, and then using the ``emcee'' MCMC sampling package \citep{ForemanMackey13} with 5000 samples discarding the first 1000 samples.
The uncertainties include accounting for the foreground uncertainties based on repeating the fitting with $\ebv_\mathrm{MW}$ plus and minus the uncertainties given in Table~\ref{tab_anc} and adding the resulting uncertainties in quadrature.

\subsection{Comparison to Previous Work}
\label{sec_prev_work}

\begin{figure}[tbp]
\epsscale{1.15}
\plotone{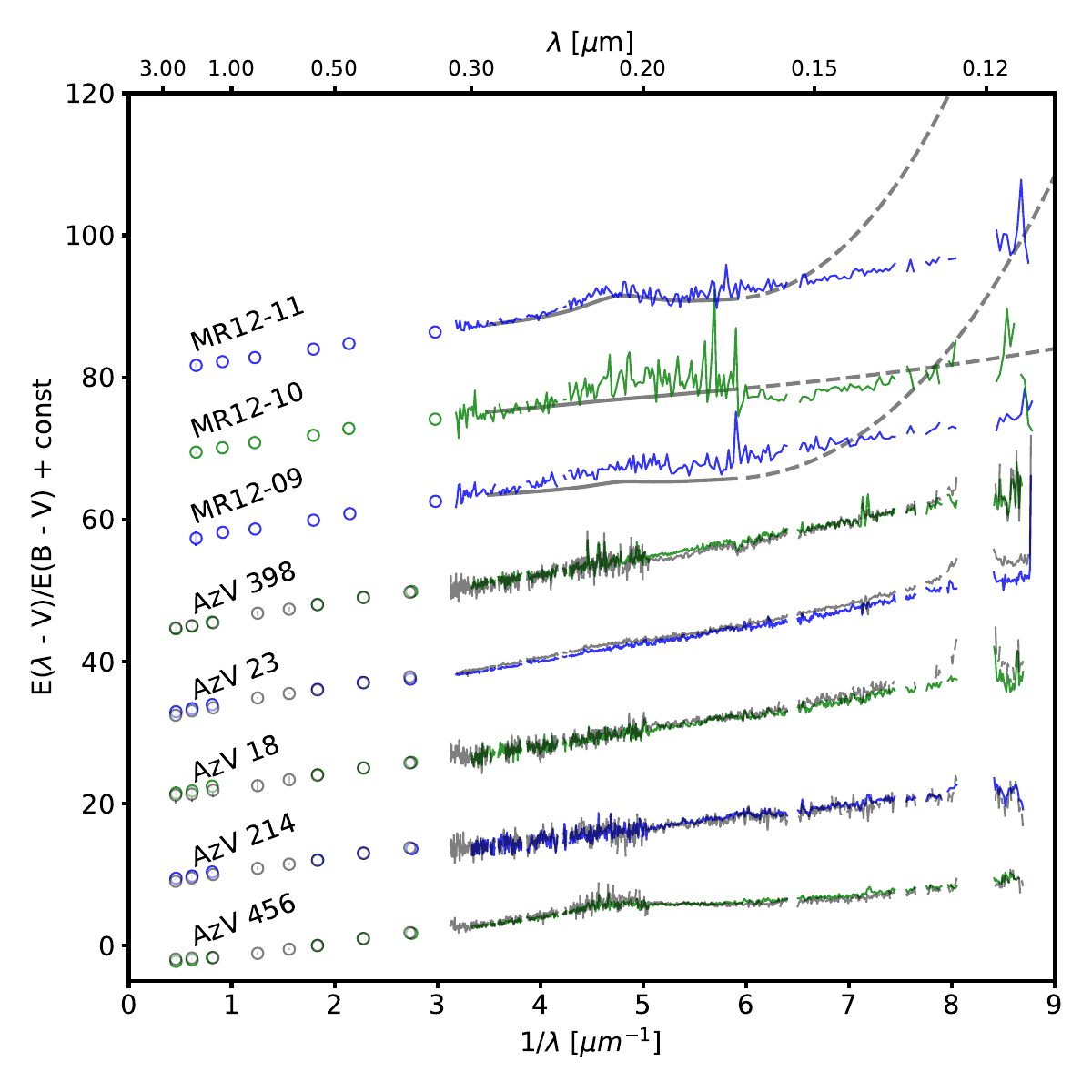}
\caption{The comparison of the MW foreground corrected (blue/green lines) and previous literature (gray lines) extinction curves are shown.
Note that some of the curves overlap so well that it is hard to see both curves.
The five AzV sightlines show the comparison with the \citet{Gordon03} sample and the 
three MR12 sightlines show the comparison with the \citet{MaizApellaniz12} sample based on their \citet{Fitzpatrick90} parameters.
For the MR12 sample, the solid lines show the region based on spectroscopy and the dashed line the region based on a single far-UV photometric measurement.
The new extinction curves agree well for the \citet{Gordon03} sample, but can be significantly different for the MR12 sample.
\label{fig_prev_curves}}
\end{figure}

Eight of the sightlines measured in this work have also been measured previously and the comparisons to them are shown in Fig.~\ref{fig_prev_curves}.
The five sightlines towards AzV stars were measured by \citet{Gordon03} using the standard pair method where the comparisons used were {\it observed} spectra of SMC stars with low reddening, implicitly correcting for MW foreground dust.
Our new and the previous curves for these five sightlines are very similar  providing validation of the foreground extinction correction for the new measurements.

The comparisons for the three sightlines towards the MR12 stars as measured by \citet{MaizApellaniz12} show significant differences.
These previous extinction curves were measured for the total MW+SMC dust column and hence for these comparisons we show our MW+SMC extinction curves.
The largest differences are seen in the FUV where the previous measurements were based on a single photometric measurement.
In the NUV where the previous measurements were based on UV prism spectroscopy, the comparisons is quite good for MR12~11, but shows significant differences for the other two stars.
For MR12~09, our measurements show a stronger linear rise with decreasing wavelength.
For MR12~10, our measurements show a marked difference, displaying a significant \fbump\ bump and far-UV curvature where the previously measured extinction curve exhibits only a linear component.
One possible explanation of the difference may be the challenges of the quite low resolution UV prism data in comparison to the new higher resolution slit spectroscopy.

\section{SMC Properties}
\label{sec_discussion}

\subsection{Environmental Dependence} 

\begin{figure*}[tbp]
\epsscale{1.2}
\plotone{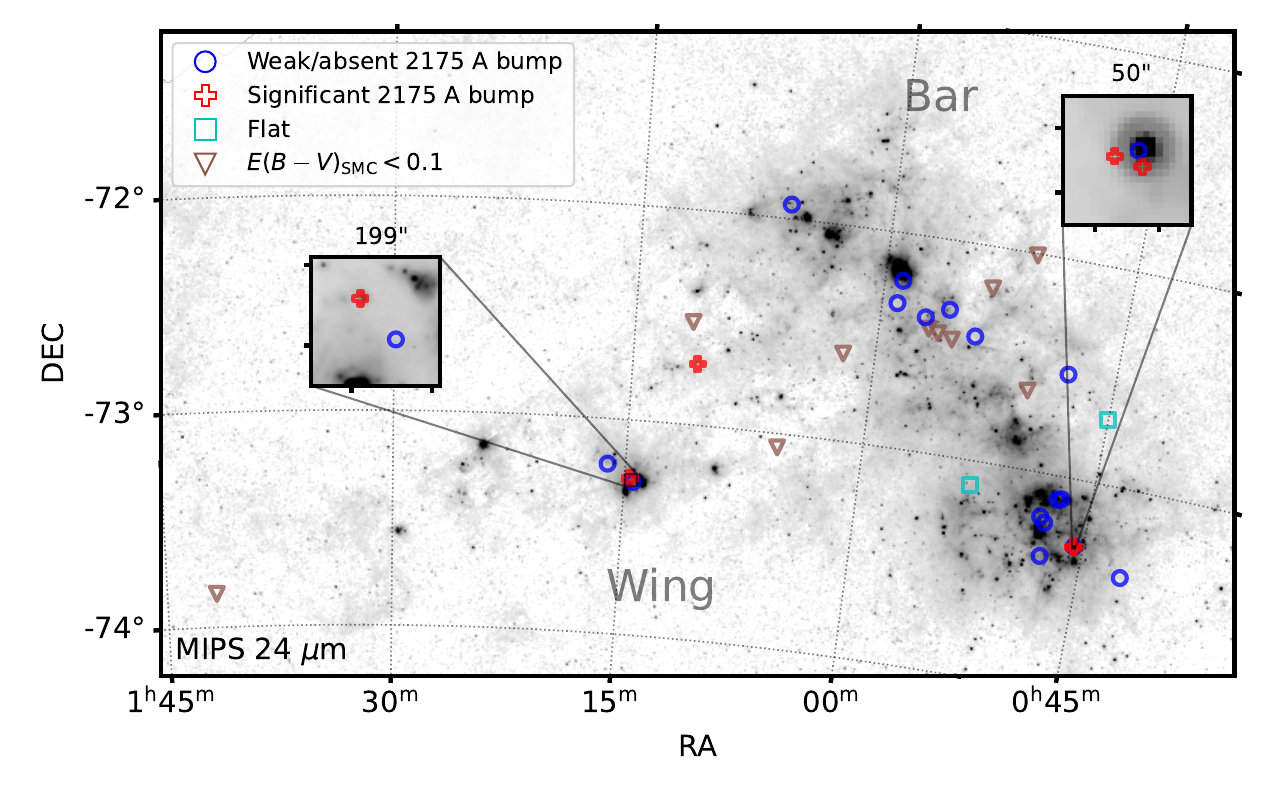}
\caption{The locations of the extinction curve sightlines are shown on a MIPS~24~\micron\ image.
MIPS 24~\micron\ traces the dust distribution with a focus on the youngest, embedded star formation.
Zoomed images of two regions containing sightlines that probe small angular scales are shown.
The different groups are shown with different symbols.
The weak/absent \fbump\ bump sightlines are located throughout the galaxy including the Bar (tilted region on the right) and Wing (more diffuse region extending from the Bar towards the left).
The sightlines with significant \fbump\ bumps are located in three well separated regions including in the Bar and Wing.
\label{fig_bump_vs_position}}
\end{figure*}

\citet{Gordon98} speculated that nearby star formation was the cause of the bumpless sightlines seen towards some SMC stars.
This was suggested by the apparent association of the star-forming SMC Bar with the bumpless sightlines and the more quiescent SMC Wing with the one then-known sightline with a bump.
Based on only four sightlines, this simple picture was challenged by the finding of two sightlines with and two sightlines without bumps very near each other ($\sim$10\arcsec) in a small region in the SMC Bar by \citep{MaizApellaniz12}.

Using our larger sample, the dependence of the dust extinction properties on environmental factors can be investigated in more detail by examining the locations of the sightlines on a representative SMC image.
This is done in Fig.~\ref{fig_bump_vs_position} where the sightline positions are superimposed on the SMC MIPS 24~\micron\ image \citep{Gordon14}, with different symbols for the four different types of curves.
The 24~\micron\ emission traces embedded star formation, and effectively represents the combination of star formation and dust.
It is clear from the image that sightlines both with and without \fbump\ bumps are found throughout the SMC, including the Bar and Wing, and have no clear general association with 24~\micron\ emission.
In addition, the two zoomed images in Fig.~\ref{fig_bump_vs_position} show that there are sightlines with and without \fbump\ bumps in close proximity as projected on the sky.
This was initially discovered by \citet{MaizApellaniz12} and strengthened by our finding of a second region in the SMC Wing showing similar spatial variations.

Given the behavior of this larger sample, it is certainly not clear whether nearby star formation is an important factor in the determination of extinction curve properties.
While the simple 2D positional analysis could be compromised by the line-of-sight depth of the SMC, it may also be that other factors play significant roles.
For example, dust self-shielding could explain why dust producing significant \fbump\ bumps can be found near regions of star formation.
Evidence that this might play a role can be seen in the behavior of LMC dust extinction, where stronger \fbump\ bumps are found near the 30~Dor star-forming region, which is associated with significant molecular material \citep{Misselt99}.

\subsection{Average Curves}
\label{sec_aves}

\begin{figure*}[tbp]
\epsscale{1.15}
\plottwo{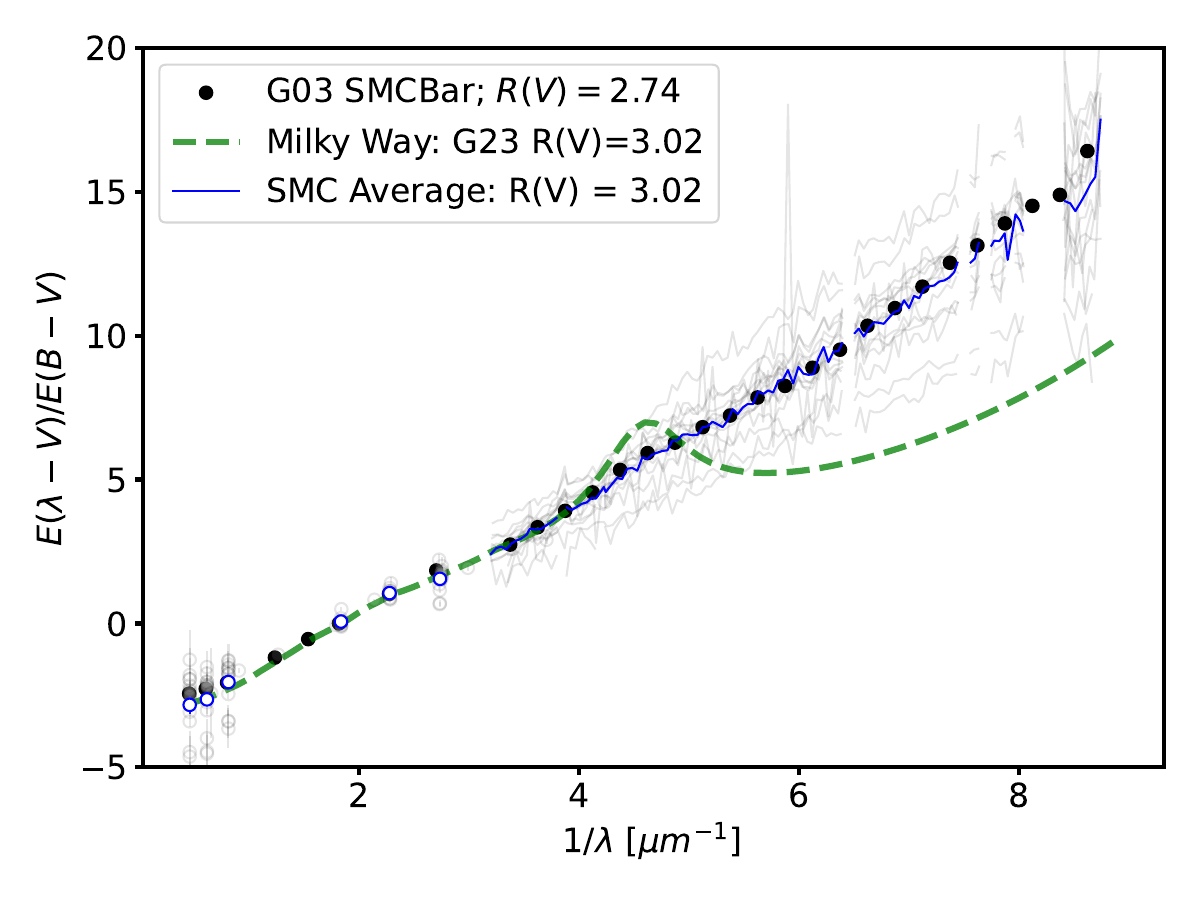}{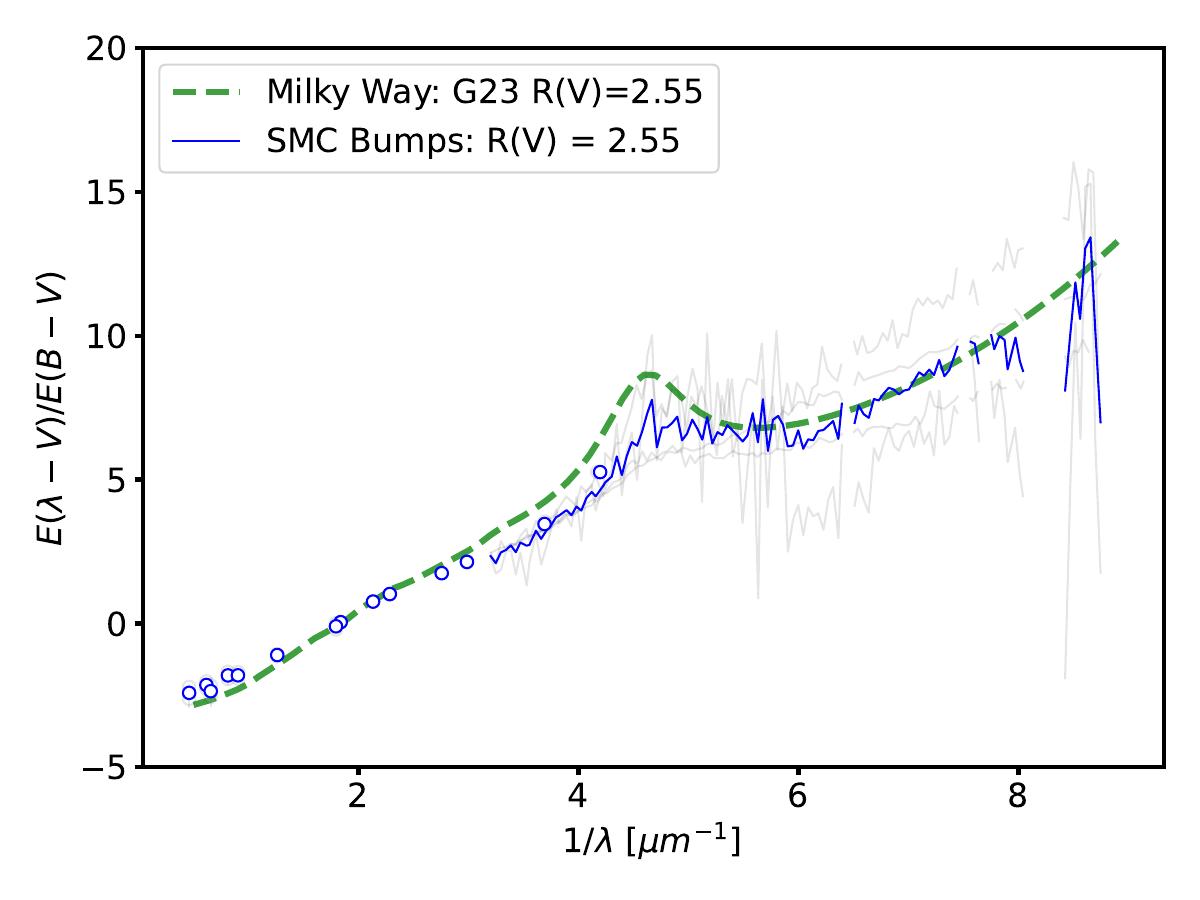}
\caption{The average extinction curves are shown for the weak/absent \fbump\ bump (aka SMC Average) and significant bumps (aka SMC Bumps) samples.
The individual curves that contributed to each average are shown in gray.
These curves have been rebinned by a factor of 10.
The averages were done with the \elvebv\ curves and the \rv\ values for these averages were computed analogous to how the \av\ values were computed (see \S\ref{sec_final_curves}).
The \citet{Gordon03} SMC average (aka SMCBar) and MW average extinction for the same \rv\ as the average \citep{Gordon23} are shown for comparison.
\label{fig_aveext}}
\end{figure*}

\begin{deluxetable}{lcccc}[htbp]
\tablecaption{Sample Averages \label{tab_aveext}}
\tablehead{\colhead{$\lambda$} & \colhead{$x$} & \colhead{Band} & \colhead{SMC Average} & \colhead{SMC Bumps} \\
 \colhead{[$\micron$]} & \colhead{[\micron$^{-1}$]} & & \colhead{$A(\lambda)/A(V)$} & \colhead{$A(\lambda)/A(V)$}}
\startdata
$0.112$ & $8.904$ & IUE/STIS & $6.929 \pm 0.203$ & $4.326 \pm 0.426$ \\
$0.117$ & $8.546$ & IUE/STIS & $5.899 \pm 0.036$ & $5.480 \pm 0.105$ \\
$0.122$ & $8.202$ & IUE/STIS & $5.377 \pm 0.180$ & $4.320 \pm 0.833$ \\
$0.127$ & $7.872$ & IUE/STIS & $5.511 \pm 0.031$ & $4.806 \pm 0.085$ \\
$0.132$ & $7.555$ & IUE/STIS & $5.173 \pm 0.036$ & $4.724 \pm 0.085$ \\
$0.138$ & $7.251$ & IUE/STIS & $4.914 \pm 0.023$ & $4.444 \pm 0.059$ \\
$0.144$ & $6.959$ & IUE/STIS & $4.654 \pm 0.021$ & $4.269 \pm 0.053$ \\
$0.150$ & $6.679$ & IUE/STIS & $4.428 \pm 0.021$ & $4.048 \pm 0.056$ \\
$0.156$ & $6.411$ & IUE/STIS & $4.167 \pm 0.024$ & $3.750 \pm 0.071$ \\
$0.163$ & $6.153$ & IUE/STIS & $3.973 \pm 0.019$ & $3.689 \pm 0.048$ \\
$0.169$ & $5.905$ & IUE/STIS & $3.804 \pm 0.023$ & $3.656 \pm 0.043$ \\
$0.176$ & $5.668$ & IUE/STIS & $3.615 \pm 0.018$ & $3.563 \pm 0.042$ \\
$0.184$ & $5.439$ & IUE/STIS & $3.435 \pm 0.016$ & $3.452 \pm 0.030$ \\
$0.192$ & $5.221$ & IUE/STIS & $3.272 \pm 0.016$ & $3.303 \pm 0.011$ \\
$0.200$ & $5.011$ & IUE/STIS & $3.162 \pm 0.015$ & $3.365 \pm 0.018$ \\
$0.208$ & $4.809$ & IUE/STIS & $3.026 \pm 0.014$ & $3.384 \pm 0.012$ \\
$0.217$ & $4.615$ & IUE/STIS & $2.894 \pm 0.014$ & $3.420 \pm 0.040$ \\
$0.226$ & $4.430$ & IUE/STIS & $2.735 \pm 0.012$ & $3.170 \pm 0.028$ \\
$0.235$ & $4.251$ & IUE/STIS & $2.571 \pm 0.015$ & $2.831 \pm 0.023$ \\
$0.245$ & $4.080$ & IUE/STIS & $2.409 \pm 0.010$ & $2.659 \pm 0.015$ \\
$0.255$ & $3.916$ & IUE/STIS & $2.328 \pm 0.011$ & $2.536 \pm 0.007$ \\
$0.266$ & $3.759$ & IUE/STIS & $2.174 \pm 0.009$ & $2.318 \pm 0.011$ \\
$0.277$ & $3.607$ & IUE/STIS & $2.091 \pm 0.010$ & $2.353 \pm 0.032$ \\
$0.289$ & $3.462$ & IUE/STIS & $1.968 \pm 0.009$ & $2.203 \pm 0.023$ \\
$0.301$ & $3.323$ & IUE/STIS & $1.888 \pm 0.010$ & $2.188 \pm 0.007$ \\
$0.363$ & $2.756$ & JohnU & $1.514 \pm 0.037$ & $1.684 \pm 0.000$ \\
$0.438$ & $2.284$ & JohnB & $1.349 \pm 0.012$ & $1.400 \pm 0.007$ \\
$0.544$ & $1.837$ & JohnV\tablenotemark{a} & $1.021 \pm 0.012$ & $1.017 \pm 0.000$ \\
$1.231$ & $0.812$ & JohnJ & $0.324 \pm 0.069$ & $0.293 \pm 0.049$ \\
$1.622$ & $0.617$ & JohnH & $0.125 \pm 0.081$ & $0.162 \pm 0.049$ \\
$2.174$ & $0.460$ & JohnK & $0.062 \pm 0.104$ & $0.055 \pm 0.081$
\enddata
\tablenotetext{a}{\kchange{The average \alav\ values at V band are not exactly 1.0 due to the use of the SED fitting based $E(55-44)$ values corrected for \ebvmw\ for the \elvebv\ normalized curves.  As discussed in \S\ref{sec_stellar_model}, $E(55-44)$ is close to but not exactly \ebv.}}
\end{deluxetable}

The averages for the two main SMC extinction samples are given in Fig.~\ref{fig_aveext} and tabulated in Table~\ref{tab_aveext}.
The \rv\ values and FM90 parameters for these two averages are given in Tables~\ref{tab_ext_col_param} and \ref{tab_ext_bump_params}.
These averages are provided as they are often used in dust grain modeling, accounting for extinction along sightlines without measured extinction curves, and comparisons to similar averages in other galaxies.

We designate the average created from the 16 sightlines in the weak/absent \fbump\ bump sample as the "SMC Average" curve.
This is motivated by our finding that the extinction curves in this sample are seen throughout the SMC and include most SMC sightlines.
The SMC Average curve is very similar to the "SMCBar" average curve from \citet{Gordon03} that was determined from only four sightlines in the SMC Bar.
The main difference is that the new larger sample has an average $\rv = 3.02 \pm 0.18$, consistent with the MW average \rv\ whereas the SMCBar sample of \citet{Gordon03} had an average $\rv$ of 2.74.

The average of the four sightlines in the significant \fbump\ bump sample is designated as the "SMC Bumps" curve.
This average has a weaker \fbump\ bump and lower extinction in the near-UV than the MW $\rv = 2.55$ average, but very similar far-UV extinction.
This average is less certain as it is composed of only four sightlines and it may change significantly when a larger sample of SMC sightlines with \fbump\ bumps is available.

\subsection{The 2175~\AA\ bump and MIR Carbonaceous Features}
\label{sec_bump_qpah}

\begin{figure*}[tbp]
\epsscale{1.15}
\plotone{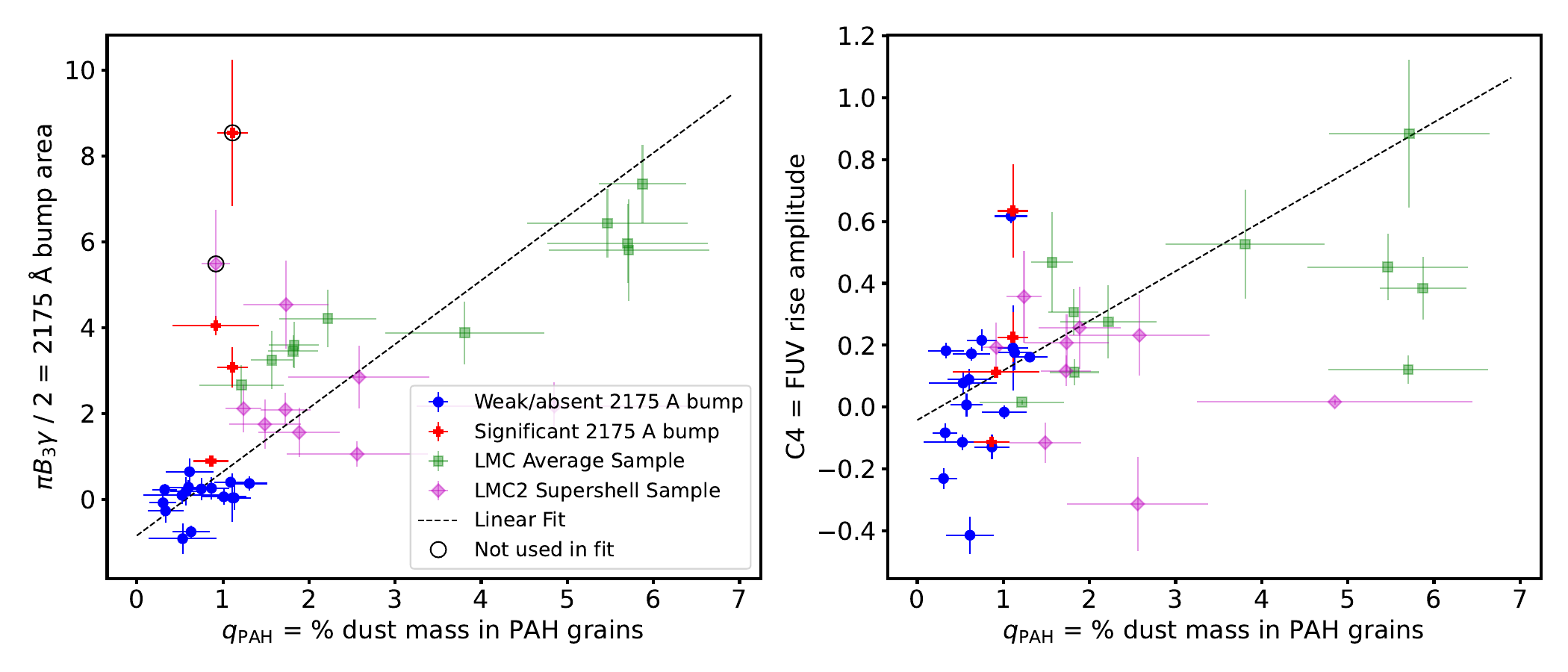}
\caption{The area of the \fbump\ bump ($\pi B_3 \gamma / 2$) and the amplitude of the FUV rise are plotted versus the percentage mass fraction of PAH grains (\qpah).
\qpah was derived by fitting the IR emission SED at sightline positions.
The \fbump\ bump and $C_4$ values for the SMC extinction sightlines are from this paper and from \citet{Misselt99} for the LMC sightlines.
The \qpah\ values for the SMC are described in \S\ref{sec_anc_data} and the LMC values obtained from the same source \citep{Chastenet19}.
The linear fit was determined using fitting with outlier removal and the points removed are indicated.
\label{fig_bump_vs_qpah}}
\end{figure*}

The \fbump\ bump is most likely due to small carbonaceous grains \citep[e.g.,][]{Mathis94} and the MIR carbonaceous features (aka PAH, UIR, AEF, etc. features) are definitely due to small carbonaceous grains \citep{Duley81}.
This has lead to the hypothesis that these two features are due to the same small carbonaceous grains \citep{Draine07model}.
This predicts that these two features should be correlated as they are from the same grain sizes and composition.
These two features are seen to be correlated in the MW for some, but not all of the MIR carbonaceous features \citep{Massa22}.
Specifically, the 6.2~\micron\ features are not correlated with the \fbump\ bump, but three other features (7.6, 8.5, and 11.3~\micron) are correlated.

We have have tested this hypothesis in the SMC (and LMC) as well as currently possible in Fig.~\ref{fig_bump_vs_qpah} where the \fbump\ bump area is plotted versus \qpah, the mass fraction of ``PAH'' grains as compared to the total mass of dust.
MW points are not included as the \qpah\ measurements for the extinction sightlines are not available.
In addition, this figures also shows the correlation with the FUV rise amplitude as it is postulated to be the wing of a carbonaceous feature peaking around $\sim$715~\AA\ \citep{Joblin92, Li01, Gordon09FUSE}.
The \qpah\ values are determined by fitting the Spitzer and Herschel observations at a resolution of 42\arcsec\ \citep{Chastenet19} and, hence, are averages around the extinction sightlines and through the entire column in the galaxy at that location.
This is in contrast to the extinction measurements that are for pencil beams along the sightline towards each star.
Therefore, the \qpah\ versus \fbump\ bump area or FUV rise correlations are unlikely to be perfect as the \qpah\ value is averaged over a larger volume than \fbump\ bump.
Note that the \qpah\ values so derived are driven by the IRAC 8~\micron\ photometry that measures the strong 7.6 and 8.5~\micron\ features that \citet{Massa22} found were correlated with the \fbump\ bump in the MW.
All measurements are by definition normalized to the total amount of dust, with \qpah\ normalized to the total mass of dust grains and extinction values are normalized to the total dust column as measured by \ebv.

Looking at just the SMC points in Fig.~\ref{fig_bump_vs_qpah}, there is no clear correlation.
For example, there is a large range of \qpah\ values for \fbump\ bump areas around zero and a large range of \fbump\ areas for \qpah\ values around 1\%.
Combining the SMC measurements with those from the LMC \citep{Misselt99, Chastenet19}, a different picture emerges where the two measurements are clearly correlated.
To illustrate this point, a linear fit with outlier removal that accounts for the significant uncertainties on both the x and y values \citep{Gordon23} was done.
The data points are scattered around the linear fit lines with a larger scatter than their uncertainties would predict.
The deviations from the linear fit could very well be explained by variations between the pencil beam extinction measurements and the 42\arcsec\ spatially averaged \qpah\ measurements.
Higher spatial resolution \qpah\ observations like those possible with JWST would test this explanation of the scatter.

\section{SMC, LMC, and MW Comparisons}
\label{sec_discussion2}

In this section, we analyze the details of the SMC extinction curves in context with equivalent measurements in the MW and LMC.
The SMC extinction curves have significant variation and how this variation compares to what is seen in these other two galaxies is expected to provide insight into dust grain evolution as has already been discussed by \citet{Gordon03}.
The previous subsection (\S\ref{sec_bump_qpah}) illustrates the strength of combining SMC and measurements from another galaxy.
It was after adding the LMC measurements that the correlation between the \fbump\ bump and \qpah\ became clear.

\subsection{General Sightline Properties}

\begin{figure*}[tbp]
\epsscale{1.20}
\plotone{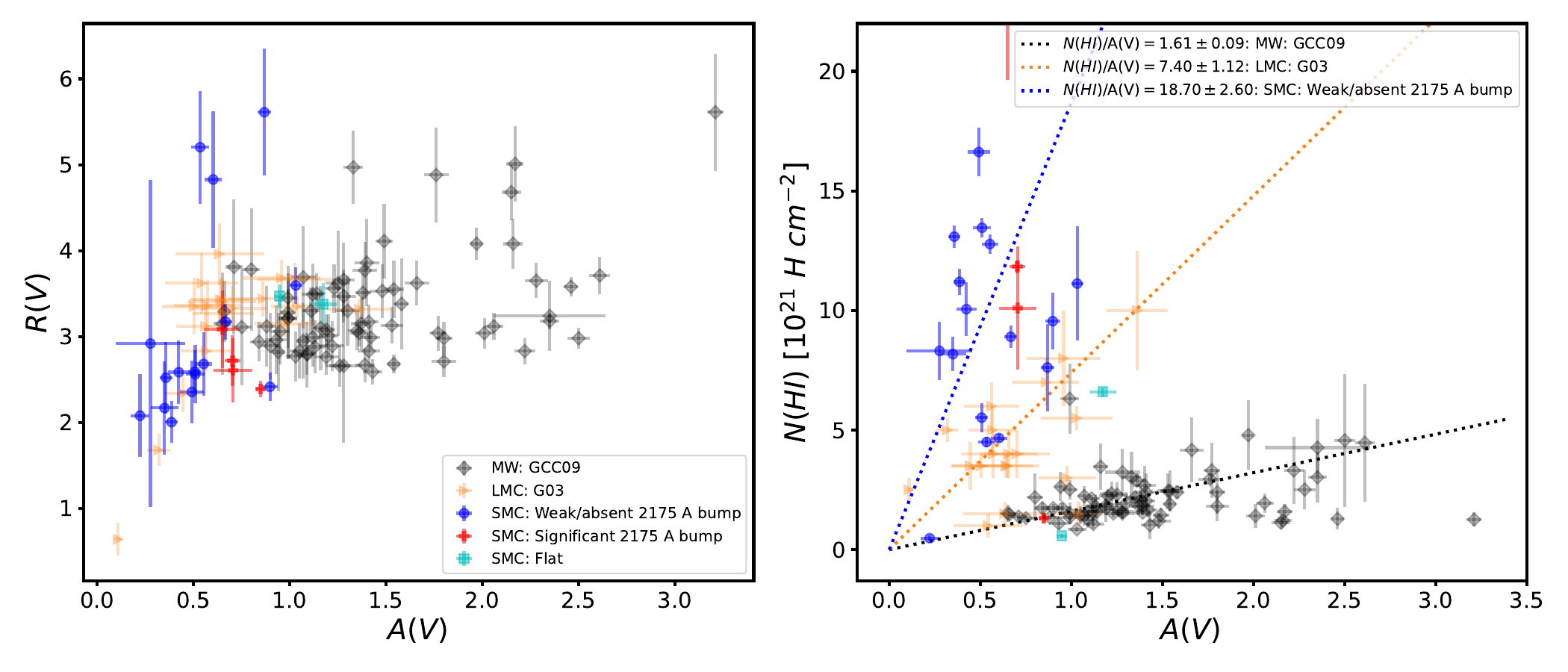}
\caption{The general sample properties are shown.
The dust grain size average measure \rv\ is plotted versus the dust column \av\ on the left.
On the right, the gas column as measured by \nhi\ is plotted versus the dust column \av\ \kchange{measured in magnitudes}.
Note that the measurement for MR12~10 is off the top of the plot and only the bottom error bar of this very uncertain measurement shown. 
In addition to the SMC subsamples from this paper, the existing LMC sample \citep{Misselt99, Gordon03}, and MW \citep{Gordon09FUSE} are shown.
The dotted lines give the average \nhiav\ for the three samples with enough measurements to be meaningful. 
\kchange{The average \nhiav\ values and uncertainties are given in the legend.}
The \citet{Gordon09FUSE} sample is shown as it is the only large MW sample with full UV extinction curves and measured \nhi\ values \citep{Gordon09FUSE, VanDePutte23}.
\label{fig_samp_props}}
\end{figure*}

While the overall impression gained from SMC extinction is that it is markedly and systematically different from that seen in the MW and LMC, nevertheless there is some overlap and some continuity in their properties.
To illustrate this, we begin by plotting in Fig.~\ref{fig_samp_props} the general properties of the 22 sightlines with $\ebvsmc > 0.1$ along with existing samples for the MW and LMC.
Note that starting here all the extinction measurements discussed are for the SMC only quantities (e.g., $\ebvsmc\ = \ebv$).

The \rv\ values for most of the SMC sightlines fall within the range of those seen in the MW and LMC.
For the gas-to-dust values as measured by \nhiav, the right panel of Fig.~\ref{fig_samp_props} shows that these values are generally higher than those found in the MW or LMC.
This is expected as the gas-to-dust ratio is a function of metallicity and the SMC is more metal poor than both the MW and LMC.
The average \nhiav\ value for the SMC of $(18.7 \pm 2.6) \times 10^{21}~\mathrm{H}~\mathrm{cm}^{-2}~\mathrm{mag}^{-1}$ is higher than the previously the measured value of $(13.2 \pm 1.0) \times 10^{21}~\mathrm{H}~\mathrm{cm}^{-2}~\mathrm{mag}^{-1}$ \citep{Gordon03} likely due to this value being based on only four sightlines.
Our average is lower than the more recent value of $(27.5 \pm 3.5) \times 10^{21}~\mathrm{H}~\mathrm{cm}^{-2}~\mathrm{mag}^{-1}$ \citep{Yanchulova21} possibly due to this value being measured in a small region of the SMC.
In addition there is a larger scatter than explained by measurement uncertainties in the individual values.

The values of \nhiav\ do not scale linearly with the ratio to the MW metallicity.
The ratio of average \nhiav\ to average $\nhiav_\mathrm{MW}$ are 12 and 5 for the SMC and LMC, yet the same ratio with metallicities are only 5 and 2 \citep{Russell92, Dominguez-Guzman22}.
It addition, is striking is that there is substantial overlap in the individual sightline \nhiav\ values between each galaxy.
This is not a function of the dust column, as all three galaxies have measured \av\ values that range between 0.5 and 1.0 and all show a range of \nhi\ values for this \av\ range.
In fact, there are sightlines in the SMC (and LMC) that have MW gas-to-dust ratios.
This clearly indicates that \nhiav\ is dependent on more than just the galaxy metallicity.

\subsection{UV Parameter Correlations}

\begin{figure*}[tbp]
\epsscale{1.2}
\plotone{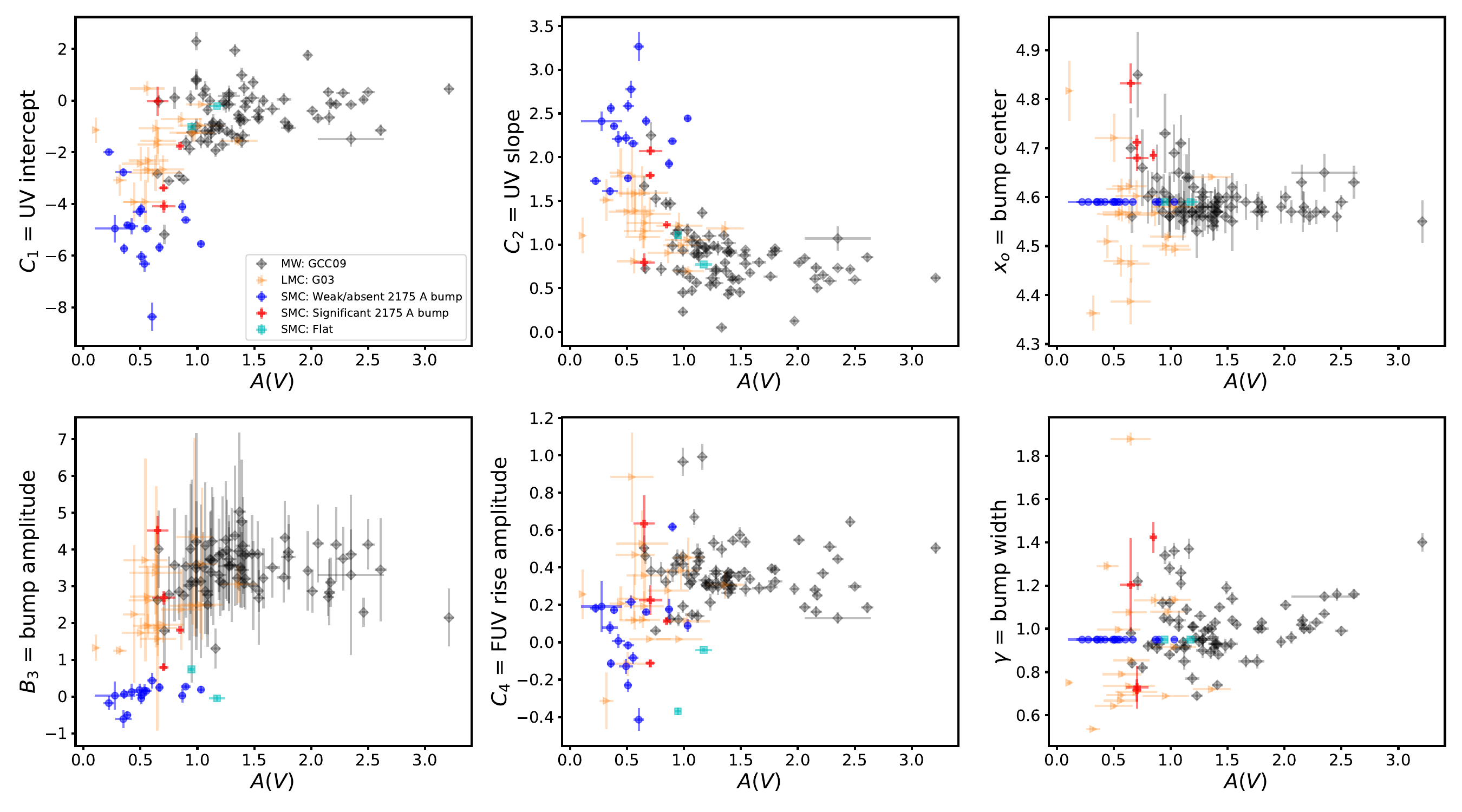}
\caption{The FM90 parameters are plotted versus dust column density as measured by \av\ \kchange{in magnitudes}.
The behavior of MW sightlines is illustrated by the \citep{Gordon09FUSE} sample.
The LMC sightlines are from \citet{Gordon03}.
The sightlines with weak/absent \fbump\ bumps used fixed values for $x_o$ and $\gamma$.
\label{fig_av_vs_fm90}}
\end{figure*}

The parameterized behavior of the SMC UV extinction curves versus dust column \av\ are plotted in Fig.~\ref{fig_av_vs_fm90} along with measurement for the LMC and MW.
These plots clearly show that the SMC (and LMC) sightlines generally probe lower dust columns than have been measured in the MW.
This is explained as MW studies generally use a cutoff of $\ebv \sim 0.2$ \citep{Valencic04, Fitzpatrick07, Gordon09FUSE}.
Lower \ebv\ values can be probed in the MW to lower dust columns \citep{Fitzpatrick05}, it is just not usually done.
The plots illustrate that the SMC sightlines generally have lower UV intercepts ($C_1$), \fbump\ bump strengths ($B_3$), and far-UV rises ($C_4$) than the MW.
The SMC sightlines have higher UV slopes ($C_2$) than the MW.
But these statements are based on the average behavior, there are clearly sightlines in the SMC (and LMC) that are indistinguishable in the plots from the MW behavior.
The four SMC sightlines with significant bumps show \fbump\ centers and widths that are shifted from the MW averages, but they are within the scatter seen in the MW that increases as \av\ decreases.
\citet{Salim20} suggested that the low \fbump\ strengths in the SMC were due to the low dust columns probed.
With the larger SMC sample, it is clear that the lack of the \fbump\ bump is not correlated with dust column given that low $B_3$ values are found even beyond $\av = 1$ where high $B_3$ values are seen in the MW.

\begin{figure*}[tbp]
\epsscale{1.15}
\plotone{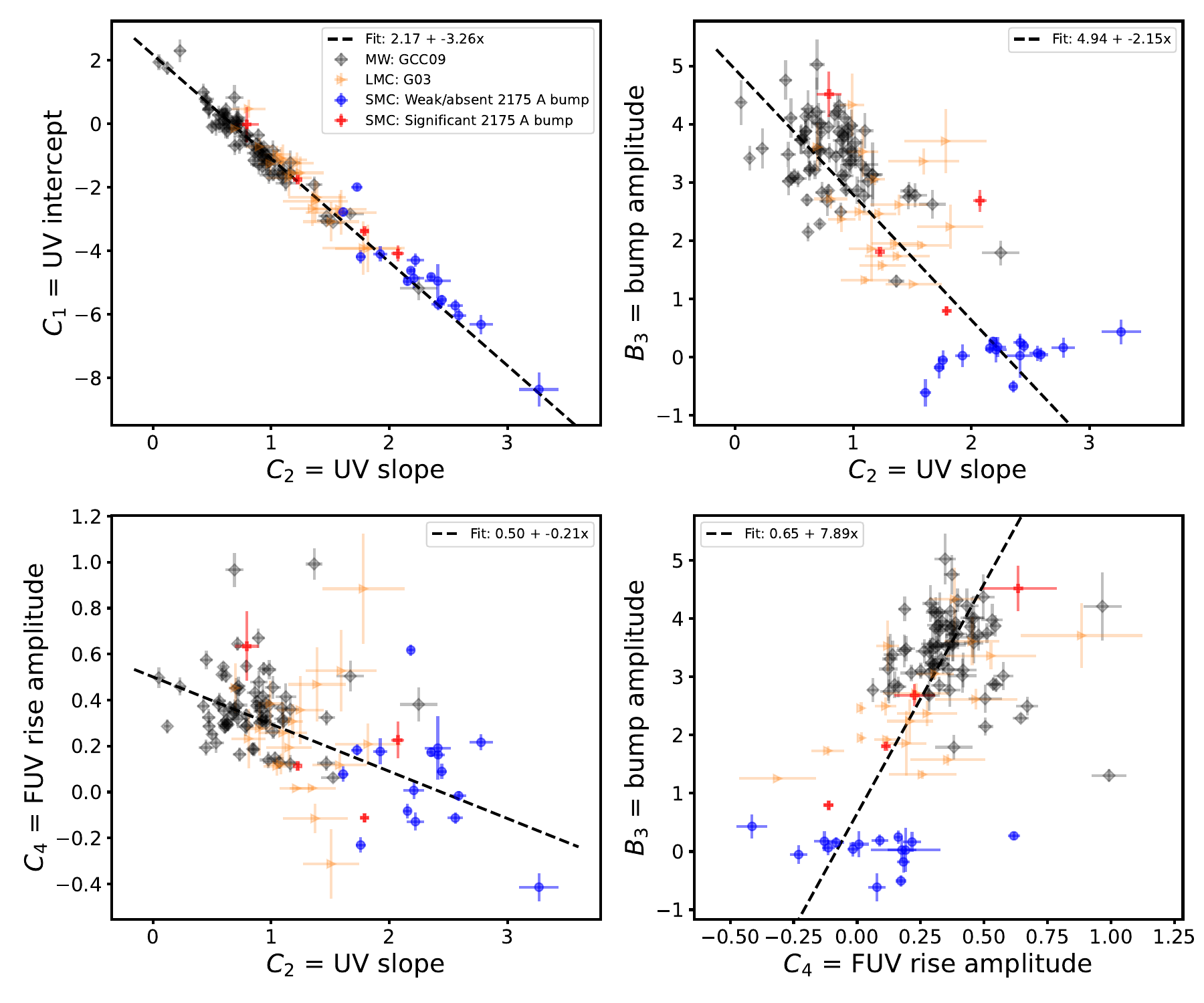}
\caption{The behavior of the FM90 parameters versus each other is shown.
The behavior of MW sightlines is illustrated by the \citep{Gordon09FUSE} sample and the LMC sightlines from \citet{Gordon03}.
Linear fits accounting for the uncertainties on both axes are shown \citep{Gordon23}.
Note that the MW and LMC $B_3$ uncertainties have been approximately reduced to account for the strong correlations between $C_3$ and $\gamma$ as they were not fit with $B_3$ explicitly and their covariances were not reported in the literature.
Only one of the three possible similar MW samples are used to avoid the fitting over-weighting the MW points in comparison to the LMC and SMC.
\label{fig_fm90_vs_fm90}}
\end{figure*}

The correlations between UV extinction parameters $C_1$, $C_2$, $B_3$, and $C_4$ are shown in Fig.~\ref{fig_fm90_vs_fm90}.
The bump center $x_o$ and width $\gamma$ are not shown as no significant correlations are seen.
The two ``flat'' SMC sightlines have been omitted to avoid biasing the fitting with curves that may be strongly impacted by systematic uncertainties.
The known strong correlation between $C_1$ and $C_2$ is seen to extend over a large range of values and the linear fit coefficients agree with and extend previous work \citep{Fitzpatrick07}.
This correlation indicates that the UV extinction curve variations pivot around a point at a specific wavelength likely due to normalizing the extinction curves by \ebv.
There are correlations seen between $C_2$, $B_3$, and $C_4$ and these are much clearer than previously seen \citep{Gordon03} due to the larger SMC sample and expanded MW sample.
Linear fits were done that account for the uncertainties on both quantities \citep{Gordon23}, but not the covariance between them as covariance is not available for the LMC and SMC samples.
Regardless, the variations seen are much larger than the uncertainties and so the correlations between parameters cannot be explained by uncertainty covariance.

The correlations indicate that as the \fbump\ bump strength weakens, far-UV rise also weakens, and the UV slope strengthens.
This is evidence that there is a family of curves with correlated changes across the UV.
This is reminiscent of the family of curves that correlates with \rv\ to describe the average behavior of MW extinction \citep{Cardelli89, Gordon23}.
Yet the family of curves indicated shows much stronger variation and is not correlated with \rv\ given the \rv\ values are similar for all the samples (see Fig.~\ref{fig_av_vs_fm90}).

\subsection{Gas-to-Dust Correlations}

\begin{figure*}[tbp]
\epsscale{1.15}
\plotone{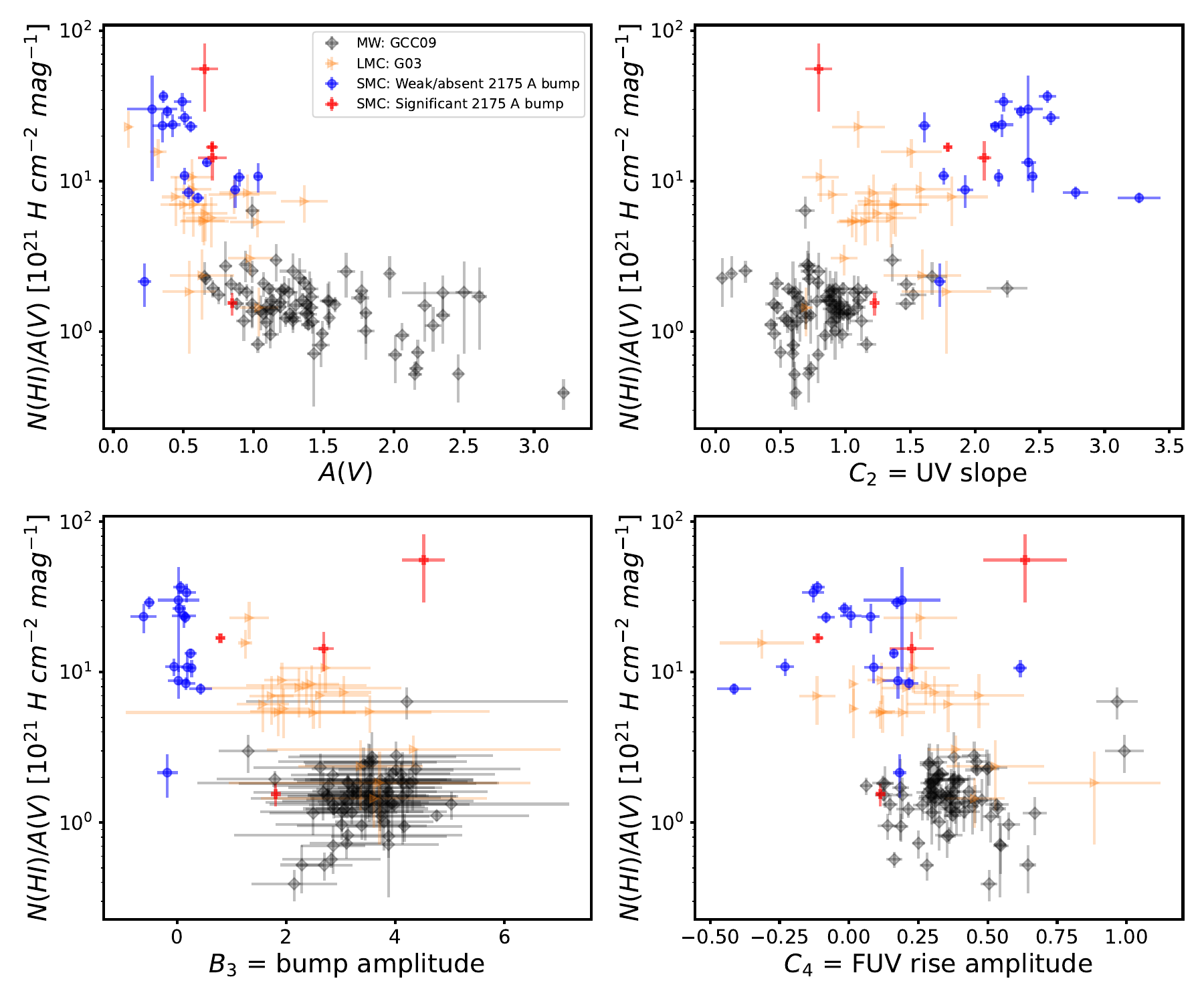}
\caption{The gas-to-dust ratio \nhiav\ is shown versus selected extinction properties.
The plot of \nhiav\ versus \av\ \kchange{(in magnitudes)} shows the general behavior of gas-to-dust increasing from the MW to the LMC to the SMC.
The plots of \nhiav\ versus $C_2$, $B_3$, and $C_4$ show clear correlations.
The plot of \nhiav\ versus the UV intercept $C_1$ is not shown as it is very similar to that for $C_2$ given the extremely strong correlation between these two parameters.
\label{fig_gdprops}}
\end{figure*}

\citet{Gordon03} speculated that the large variations in the FM90 parameters seen may be correlated with gas-to-dust as measured by \nhiav.
Fig.~\ref{fig_gdprops} plots \nhiav\ versus $C_2$, $B_3$, and $C_4$ as well as \av.
The \nhiav\ versus \av\ plot shows that while the average gas-to-dust value decreases from the SMC to the LMC to the MW, there are sightlines in each galaxy that have \nhiav\ ratios that are like one or both of the other galaxies.
In particular, the SMC shows that some sightlines have \nhiav\ values like those seen in the LMC and MW.
From the other panels it is seen that the UV slope $C_2$ and \fbump\ bump amplitude $B_3$ are clearly correlated with \nhiav.
The correlation with the FUV rise amplitude $C_4$ is present, but significantly weaker.

The correlations of the UV parameters with \nhiav\ indicates that the extinction curve shape variations may be traced by variations in the gas-to-dust ratio.
The gas-to-dust ratio varies with metallicity as dust is composed almost entirely of metals (i.e., carbonaceous and silicate grains) and the impact of accretion/ablation of gas phase atoms onto and off of dust grains.
The metallicity is not seen to vary significantly across the LMC or SMC \citep{Russell92, Dominguez-Guzman22} and neither would the \nhiav\ ratio predicted by metallicity alone.
Thus variations in the \nhiav\ ratio are most likely due to accretion and/or ablation of dust grains.
The SMC (and LMC) show variations of up to a factor of 10 in the \nhiav\ ratios, while the MW shows a much smaller variation.
The smaller MW variation may be due to the fact we observe from inside the MW and hence the dust extinction probes a relatively small region 1--2~kpc around the Sun \citep{Valencic04}.
While for the LMC and SMC, we probe sightlines across the face of both galaxies.
However, the SMC in particular has considerable line-of-sight depth ($\sim 20\rm\,kpc$), and may be composed of distinct structures \citep{murray24}, and therefore the position of each UV-bright star along the line of sight will affect the interpretation of its \nhiav\ ratio.

A full investigation of the variations of extinction with \nhiav\ is beyond the scope of this paper.
Such an investigation must also include a simultaneous accounting for the known MW extinction variations that are strongly correlated with \rv\ \citep{Cardelli89, Gordon23}.
The potential exists for a two parameter relationship based on \rv\ and \nhiav\ extending the existing MW \rv\ dependent extinction relationship that would explain the variations seen in the MW, LMC, and SMC.
Investigating if an \rv\ and \nhiav\ extinction relationship is possible will be presented in a followup paper, including a complete accounting for the full covariance and outliers.
That such a two parameter relationship is possible, is clear given that it has already been proposed based on the \fbump\ bump amplitude \citep{Zagury07} or an arbitrary $f_A$ mixture coefficient \citep{Gordon16}.
Our finding that the variations are correlated with the gas-to-dust ratio provides a physical basis that is not a specific property of the extinction curve itself (i.e., \fbump\ bump amplitude) or an arbitrary parameter (i.e., $f_A$).

\section{Summary}
\label{sec_conclusions}

We have presented a sample of UV extinction curves for the SMC based on UV spectra from the IUE archive and three HST/STIS programs.
This sample expands the number of sightlines with $\ebvsmc > 0.1$ and UV spectroscopic measurements including \nhi\ from Ly$\alpha$ from five to 22.
All the extinction curves have been determined by comparison to stellar atmosphere models and corrected for MW foreground extinction.
Of these sightlines, 16 are seen that are strongly rising to shorter wavelengths with no significant \fbump\ bump, four have significant \fbump\ bumps, and two are relatively flat in the UV with no significant \fbump\ bumps.

The sightlines with and without \fbump\ bumps are both distributed throughout the SMC.
There are two regions that show strong variation in extinction curve shape over small spatial scales.
This confirms and expands the results seen in the combination of the two previous studies \citep{Gordon03, MaizApellaniz12}.

We present the SMC Average and SMC Bumps averages based on the weak/absent and significant \fbump\ bump samples, respectively.
The SMC Average curve is based on 16 sightlines and is extremely similar that presented for the SMC Bar based on four sightlines by \citep{Gordon03}.
Interestingly, the SMC Average $\rv = 3.13$ very similar to the MW average \rv\ yet the UV extinction curve is very different.
The SMC Bumps curve shows a weaker \fbump\ than the MW equivalent curve for the SMC Bumps $\rv = 2.59$.
These averages illustrate that the majority of the SMC curves are not explained by the MW \rv\ dependent relationship as shown in this paper and previously by \citet{Gordon03}.

The hypothesis that the \fbump\ bump and mid-IR carbonaceous (aka PAH) features is tested by correlating the \fbump\ bump area with the mass fraction of ``PAH grains'' \qpah.
We find that these two measurements are correlated in the combined SMC and LMC samples supporting this hypothesis.
This extends the results from the MW \citep{Massa22} to lower metallicities and \qpah\ values.
In addition, we find a suggestive correlation of \qpah\ with the FUV rise amplitude indicating this feature may also be due to the same grains.

The UV extinction as parameterized by the UV intercept $C_1$, UV slope $C_2$, amplitude of the \fbump\ bump $B_3$, and far-UV rise amplitude $C_4$ are seen in general to have values that are quite different from those seen in the LMC and MW.
Yet, there are SMC sightlines that have such parameters that easily fall with those seen in the LMC and MW.
These parameters are seen to be strongly correlated with each other in the combined SMC, LMC, and MW results indicating that as the UV slope becomes steeper, the \fbump\ bump and the far-UV rise becomes weaker.
This indicates that there is a family of curves that likely explains the majority of the variations seen.
We find that the UV parameters are roughly correlated with gas-to-dust as measured by \nhiav\ confirming the tentative suggestion of \citet{Gordon03}.
Thus, it is possible that the general behavior of dust extinction in the SMC, LMC, and MW may be explained by two parameters.
These two parameters could be \nhiav\ as shown here and \rv\ as has been shown in the MW \citep{Cardelli89, Gordon23}.
Further investigations of a two parameter relationship will be presented in a followup paper.

The code used for the analysis and plots is available\footnote{\url{https://github.com/karllark/hst_smc_ext}}
\footnote{\url{https://github.com/karllark/measure_extinction}}
\footnote{\url{https://github.com/karllark/extinction_ensemble_props}} \citep{hstsmcext, measureextinction, extensembleprops}. 
The STIS data used in this paper can be found in MAST: \dataset[10.17909/qw8x-9y41]{http://dx.doi.org/10.17909/qw8x-9y41}.
The extinction curves measured and analyzed are available \citep{smcdata}.
The averages are available as the G24\_SMCAvg and G24\_SMCBumps average models in the dust\_extinction package\footnote{\url{https://github.com/karllark/dust_extinction}}  \citep{dustextinction}.

\begin{acknowledgements}
We thank Petia Yanchulova Merica-Jones for providing the SMIDGE photometry for the MR12 stars.
This paper benefited from discussions of preliminary results with the ISM*@ST group\footnote{\url{https://www.ismstar.space/}}. 
This research is based on observations made with the NASA/ESA Hubble Space Telescope obtained from the Space Telescope Science Institute, which is operated by the Association of Universities for Research in Astronomy, Inc., under NASA contract NAS 5–26555. These observations are associated with programs 8198, 12258, and 14225.
\end{acknowledgements}

\facilities{IUE, HST (STIS)}

\software{Astropy \citep{astropy:2013, astropy:2018, astropy:2022}; dust\_extinction \citep{dustextinction}; measure\_extinction \citep{measureextinction}}

\typeout{}
\bibliography{dust, exgal, stars, instr, mc}{}
\bibliographystyle{aasjournal}

\appendix

\section{Photometry}
\label{data_photometry}

\begin{deluxetable*}{lcccccc} 
\tabletypesize{\footnotesize}
\tablecaption{Sample UBV Photometry} \label{tab_UBV}
\tablehead{ 
\colhead{Name}                           &
\colhead{$U$}                           & 
\colhead{$B$}                           &
\colhead{$V$}                           &
\colhead{$B-V$}                         & 
\colhead{$U-B$}                         & 
\colhead{Ref}}
\startdata
2DFS 413 & \nodata & $16.725\pm 0.040$ & $16.715\pm 0.061$ & \nodata & \nodata & 1 \\
2DFS 626 & $15.926\pm 0.030$ & $16.651\pm 0.030$ & $16.483\pm 0.031$ & \nodata & \nodata & 1 \\
2DFS 662 & $14.120\pm 0.031$ & $14.962\pm 0.021$ & $14.891\pm 0.049$ & \nodata & \nodata & 1 \\
2DFS 699 & $13.465\pm 0.030$ & $14.141\pm 0.023$ & $14.063\pm 0.030$ & \nodata & \nodata & 1 \\
2DFS 3014 & $15.878\pm 0.035$ & $16.633\pm 0.035$ & $16.637\pm 0.033$ & \nodata & \nodata & 1 \\
2DFS 3030 & $14.143\pm 0.033$ & $15.044\pm 0.029$ & $15.045\pm 0.022$ & \nodata & \nodata & 1 \\
2DFS 3171 & $14.987\pm 0.054$ & $15.790\pm 0.034$ & $15.635\pm 0.024$ & \nodata & \nodata & 1 \\
AzV 4 & \nodata & \nodata & $13.835\pm 0.033$ & $ \phm{-}0.094\pm 0.005$ & $-0.757\pm 0.010$ & 2 \\
AzV 18 & \nodata & \nodata & $12.420\pm 0.044$ & $ \phm{-}0.041\pm 0.006$ & $-0.794\pm 0.021$ & 3 \\
AzV 23 & \nodata & \nodata & $12.244\pm 0.004$ & $ \phm{-}0.084\pm 0.002$ & $-0.672\pm 0.008$ & 3 \\
AzV 70 & \nodata & \nodata & $12.413\pm 0.013$ & $-0.154\pm 0.013$ & $-1.003\pm 0.016$ & 3 \\
AzV 86 & \nodata & \nodata & $12.800\pm 0.035$ & $-0.147\pm 0.015$ & $-0.966\pm 0.005$ & 2 \\
AzV 132 & \nodata & \nodata & $13.630\pm 0.020$ & $ \phm{-}0.090\pm 0.015$ & $-0.750\pm 0.025$ & 2 \\
AzV 214 & \nodata & \nodata & $13.416\pm 0.013$ & $ \phm{-}0.038\pm 0.007$ & $-0.803\pm 0.007$ & 3 \\
AzV 218 & $12.643\pm 0.030$ & $13.554\pm 0.030$ & $13.636\pm 0.030$ & \nodata & \nodata & 1 \\
AzV 289 & \nodata & \nodata & $12.396\pm 0.026$ & $-0.118\pm 0.009$ & $-0.984\pm 0.013$ & 3 \\
AzV 380 & \nodata & \nodata & $13.534\pm 0.007$ & $-0.109\pm 0.010$ & $-0.918\pm 0.009$ & 3 \\
AzV 398 & \nodata & \nodata & $13.889\pm 0.026$ & $ \phm{-}0.100\pm 0.022$ & $-0.820\pm 0.021$ & 3 \\
AzV 456 & \nodata & \nodata & $12.888\pm 0.019$ & $ \phm{-}0.109\pm 0.009$ & $-0.785\pm 0.015$ & 3 \\
AzV 462 & \nodata & \nodata & $12.566\pm 0.017$ & $-0.126\pm 0.012$ & $-0.914\pm 0.014$ & 1 \\
BBB 280 & \nodata & \nodata & $14.480\pm 0.020$ & $-0.020\pm 0.015$ & $-0.790\pm 0.025$ & 4 \\
NGC330 ELS 110 & \nodata & $16.330\pm 0.021$ & $16.381\pm 0.033$ & \nodata & \nodata & 1 \\
NGC330 ELS 114 & \nodata & $16.439\pm 0.023$ & $16.491\pm 0.030$ & \nodata & \nodata & 1 \\
NGC330 ELS 116 & \nodata & $16.429\pm 0.087$ & $16.554\pm 0.062$ & \nodata & \nodata & 1 \\
SK 191 & \nodata & \nodata & $11.860\pm 0.020$ & $-0.040\pm 0.015$ & $-0.850\pm 0.025$ & 5 \\
SMC5-398 & \nodata & $14.190\pm 0.032$ & $14.221\pm 0.057$ & \nodata & \nodata & 1 \\
SMC5-3739 & \nodata & \nodata & $14.180\pm 0.020$ & $-0.080\pm 0.015$ & $-0.470\pm 0.025$ & 2 \\
SMC5-79264 & \nodata & $15.297\pm 0.016$ & $15.192\pm 0.028$ & \nodata & \nodata & 1 \\
SMC5-82923 & $13.688\pm 0.030$ & $14.205\pm 0.030$ & $14.251\pm 0.037$ & \nodata & \nodata & 1 \\
\enddata
\tablerefs{(1) \citet{Zaritsky04LMC}, (2) \citet{Mermilliod97}, (3) \citet{Gordon03}, (4) \citet{Basinski67}, (5) \citet{Ardeberg77}}
\end{deluxetable*}

\begin{deluxetable*}{lccccccc} 
\tabletypesize{\footnotesize}
\tablecaption{{Sample \it griz {\rm and} JHK$_S$} \label{tab_grizJHK} Photometry}
\tablehead{ 
\colhead{Star}         &
\colhead{$g$}                           & 
\colhead{$r$}                           &
\colhead{$i$}                           &
\colhead{$z$}                           & 
\colhead{$J$}                           &
\colhead{$H$}                           & 
\colhead{$K_S$}        }
\startdata
2DFS 413 & $16.663\pm 0.044$ & $16.928\pm 0.039$ & $17.106\pm 0.062$ & $17.298\pm 0.020$ & $ 16.64\pm  0.02$ & $ 16.57\pm  0.04$ & $ 16.68\pm  0.13$ \\
2DFS 626 & $16.553\pm 0.020$ & $16.717\pm 0.022$ & $16.851\pm 0.054$ & $17.061\pm 0.020$ & $ 16.51\pm  0.03$ & $ 16.46\pm  0.03$ & $ 16.57\pm  0.14$ \\
2DFS 662 & $14.867\pm 0.020$ & $15.117\pm 0.020$ & $15.288\pm 0.020$ & $15.541\pm 0.020$ & $ 15.00\pm  0.01$ & $ 15.01\pm  0.02$ & $ 15.04\pm  0.04$ \\
2DFS 699 & \nodata & \nodata & \nodata & \nodata & $ 13.98\pm  0.01$ & $ 13.93\pm  0.01$ & $ 13.91\pm  0.01$ \\
2DFS 3014 & $16.558\pm 0.079$ & $16.771\pm 0.020$ & $16.929\pm 0.020$ & $17.145\pm 0.020$ & $ 16.75\pm  0.02$ & $ 16.68\pm  0.05$ & $ 16.99\pm  0.21$ \\
2DFS 3030 & $14.913\pm 0.029$ & $15.166\pm 0.020$ & $15.320\pm 0.027$ & $15.554\pm 0.035$ & $ 15.12\pm  0.01$ & $ 15.17\pm  0.02$ & $ 15.15\pm  0.04$ \\
2DFS 3171 & $15.621\pm 0.023$ & $15.747\pm 0.035$ & $15.840\pm 0.027$ & $16.012\pm 0.033$ & $ 15.49\pm  0.02$ & $ 15.52\pm  0.02$ & $ 15.54\pm  0.05$ \\
AzV 4 & ... & \nodata & \nodata & \nodata & $ 13.59\pm  0.02$ & $ 13.51\pm  0.01$ & $ 13.36\pm  0.02$ \\
AzV 18 & \nodata & \nodata & \nodata & \nodata & $ 12.40\pm  0.02$ & $ 12.35\pm  0.01$ & $ 12.33\pm  0.01$ \\
AzV 23 & \nodata & \nodata & \nodata & \nodata & $ 12.04\pm  0.01$ & $ 11.98\pm  0.01$ & $ 11.91\pm  0.01$ \\
AzV 70 & \nodata & \nodata & \nodata & \nodata & $ 12.79\pm  0.01$ & $ 12.81\pm  0.01$ & $ 12.85\pm  0.01$ \\
AzV 86 & \nodata & \nodata & \nodata & \nodata & $ 13.05\pm  0.01$ & $ 13.11\pm  0.01$ & $ 13.13\pm  0.01$ \\
AzV 132 & \nodata & \nodata & \nodata & \nodata & $13.609\pm 0.034$ & $13.223\pm 0.037$ & $13.120\pm 0.044$ \\
AzV 214 & \nodata & \nodata & \nodata & \nodata & $ 13.41\pm  0.01$ & $ 13.38\pm  0.01$ & $ 13.36\pm  0.01$ \\
AzV 218 & \nodata & \nodata & \nodata & \nodata & $ 13.87\pm  0.01$ & $ 13.92\pm  0.01$ & $ 13.93\pm  0.01$ \\
AzV 289 & \nodata & \nodata & \nodata & \nodata & $ 12.68\pm  0.01$ & $ 12.72\pm  0.01$ & $ 12.74\pm  0.02$ \\
AzV 380 & \nodata & \nodata & \nodata & \nodata & $ 13.82\pm  0.02$ & $ 13.86\pm  0.01$ & $ 13.91\pm  0.02$ \\
AzV 398 & \nodata & \nodata & \nodata & \nodata & $ 13.62\pm  0.01$ & $ 13.57\pm  0.01$ & $ 13.52\pm  0.02$ \\
AzV 456 & \nodata & \nodata & \nodata & \nodata & $ 12.82\pm  0.01$ & $ 12.83\pm  0.01$ & $ 12.83\pm  0.02$ \\
AzV 462 & \nodata & \nodata & \nodata & \nodata & $ 12.90\pm  0.02$ & $ 12.94\pm  0.01$ & $ 12.94\pm  0.02$ \\
BBB 280 & \nodata & \nodata & \nodata & \nodata & $ 14.26\pm  0.02$ & $ 14.17\pm  0.01$ & $ 14.02\pm  0.01$ \\
MR12 09 & $18.252\pm 0.029$ & $18.544\pm 0.044$ & $18.668\pm 0.072$ & $18.902\pm 0.020$ & $18.424\pm 0.066$ & $18.378\pm 0.068$ & $18.307\pm 0.081$ \\
MR12 11 & $18.513\pm 0.027$ & $18.668\pm 0.041$ & $18.721\pm 0.032$ & $18.925\pm 0.020$ & $18.220\pm 0.067$ & $18.400\pm 0.069$ & $18.426\pm 0.083$ \\
NGC330 ELS 110 & $16.265\pm 0.020$ & $16.738\pm 0.078$ & $16.957\pm 0.170$ & $17.266\pm 0.020$ & $ 16.96\pm  0.02$ & $ 16.98\pm  0.03$ & $ 17.23\pm  0.11$ \\
NGC330 ELS 114 & $16.312\pm 0.035$ & $16.748\pm 0.053$ & $16.936\pm 0.216$ & $17.237\pm 0.020$ & $ 16.92\pm  0.02$ & $ 16.98\pm  0.05$ & $ 16.93\pm  0.16$ \\
NGC330 ELS 116 & $16.328\pm 0.021$ & $16.787\pm 0.060$ & $17.011\pm 0.184$ & $17.266\pm 0.038$ & $ 16.91\pm  0.02$ & $ 16.92\pm  0.07$ & $ 16.94\pm  0.13$ \\
SK 191 & \nodata & \nodata & \nodata & \nodata & $ 12.04\pm  0.02$ & $ 12.02\pm  0.01$ & $ 12.02\pm  0.02$ \\
SMC5-000398 & $14.053\pm 0.136$ & $14.426\pm 0.020$ & \nodata & $14.784\pm 0.110$ & $ 14.38\pm  0.01$ & $ 14.42\pm  0.01$ & $ 14.38\pm  0.02$ \\
SMC5-003739 & \nodata & \nodata & \nodata & \nodata & $ 14.13\pm  0.01$ & $ 14.12\pm  0.01$ & $ 14.15\pm  0.02$ \\
SMC5-079264 & $15.182\pm 0.020$ & $15.410\pm 0.100$ & $15.557\pm 0.020$ & $15.694\pm 0.384$ & $ 15.20\pm  0.02$ & $ 15.17\pm  0.02$ & $ 15.17\pm  0.04$ \\
SMC5-082923 & $14.192\pm 0.020$ & $14.490\pm 0.020$ & \nodata & $14.892\pm 0.020$ & $ 14.46\pm  0.01$ & $ 14.45\pm  0.02$ & $ 14.51\pm  0.02$ \\
\enddata
\end{deluxetable*}

\begin{deluxetable}{lccc}
\tablecaption{Sample HST photometry \label{tab_hst}}
\tablehead{\colhead{Band} & \colhead{MR12 09} & \colhead{MR12 10} & \colhead{MR12 11}}
\startdata
F225W & $17.362 \pm 0.005$ & $18.803 \pm 0.009$ & $17.966 \pm 0.006$ \\
F275W & $17.378 \pm 0.004$ & $18.648 \pm 0.009$ & $17.713 \pm 0.004$ \\
F336W & $17.607 \pm 0.002$ & $18.643 \pm 0.004$ & $17.784 \pm 0.003$ \\
F475W & $18.410 \pm 0.002$ & $19.160 \pm 0.001$ & $18.623 \pm 0.001$ \\
F550M & $18.391 \pm 0.002$ & $19.005 \pm 0.003$ & $18.522 \pm 0.002$ \\
F814W & $18.385 \pm 0.001$ & $18.880 \pm 0.001$ & $18.376 \pm 0.001$ \\
F110W & $18.480 \pm 0.003$ & $18.828 \pm 0.002$ & $18.410 \pm 0.002$ \\
F160W & $18.443 \pm 0.002$ & $18.740 \pm 0.003$ & $18.353 \pm 0.002$
\enddata
\end{deluxetable}

The literature optical/near-infrared photometry for all the stars in our sample is given in Tables~\ref{tab_UBV}--\ref{tab_hst}.
The optical {\it UBV} photometry is given in Table~\ref{tab_UBV} in the form reported in literature where roughly half reported colors and half reported magnitudes in all three bands.
In Table~\ref{tab_grizJHK} the {\it griz} values are from the SMASH survey \citep{Nidever17} and the {\it JHK$_S$} values are from the IRSF survey \citep{Kato07}.
One exception is for AzV~132 where the photometry is from 2MASS \citep{Cutri032MASS}.
Given the faintness of the MR12 stars, HST photometry from the HST SMIDGE catalog \citet{YanchulovaMerica-Jones17, YanchulovaMerica-Jones21}.
The SMIDGE region covered a fairly small portion of the SMC, fortunately including the entire MR12 region.

\end{document}